\documentclass[usenatbib]{mnras}

\usepackage{newtxtext}

\usepackage[T1]{fontenc}
\usepackage{ae,aecompl}

\usepackage{graphicx}	
\usepackage{amsmath}	
\usepackage{amssymb}	
\usepackage{gensymb}    
\usepackage{lscape}
\usepackage{upgreek} 

\title[Variable radio sources and VLBI]{Astrometry of variable compact radio sources: A search for Galactic black hole X-ray binaries}

\author[P. Atri et al.]
{P. Atri,$^{1, 2}$\thanks{Email: atri@astron.nl}
J. C. A.~Miller-Jones,$^{2}$
A. Bahramian,$^{2}$
R. M. Plotkin,$^{3}$
T. J. Maccarone,$^{4}$
\newauthor 
B. Marcote,$^{5}$
C. O. Heinke,$^{6}$
G. R. Sivakoff,$^{6}$
A. Ginsburg,$^{7}$
J. Strader$^{8}$ and
L. Chomiuk$^{8}$
\\ \\\\ \\
$^{1}$ASTRON, Netherlands Institute for Radio Astronomy, Oude Hoogeveensedijk 4, 7991 PD Dwingeloo, The Netherlands\\
$^{2}$International Centre for Radio Astronomy Research, Curtin University, GPO Box U1987, Perth, WA 6845, Australia\\
$^{3}$Department of Physics, University of Nevada, Reno, NV 89557, USA\\
$^{4}$Department of Physics, Box 41051, Science Building, Texas Tech University, Lubbock, TX 79409-1051, USA\\
$^{5}$Joint Institute for VLBI ERIC, Oude Hoogeveensedijk 4, 7991 PD Dwingeloo, The Netherlands\\
$^{6}$Dept. of Physics, University of Alberta, CCIS 4-181, Edmonton, AB T6G 2E1, Canada\\
$^{7}$Department of Astronomy, University of Florida, PO Box
112055, USA\\
$^{8}$Center for Data Intensive and Time Domain Astronomy, Department of Physics and Astronomy, Michigan State University, East
Lansing, MI 48824, USA
}

\date{Accepted XXX. Received YYY; in original form ZZZ}

\pubyear{2021}

\hypersetup{draft}

\begin{document}
\label{firstpage}
\pagerange{\pageref{firstpage}--\pageref{lastpage}}
\maketitle

\begin{abstract}
We use the Very Long Baseline Array to conduct high precision astrometry of a sample of 33 compact, flat spectrum, variable radio sources in the direction of the Galactic plane \citep{Becker2010}. Although \citet{Becker2010} ruled out a few potential scenarios for the origin of the radio emission, the study could not rule out that these sources were black hole X-ray binaries (BHXBs). Most known BHXBs are first detected by X-ray or optical emission when they go into an outburst, leaving the larger quiescent BHXB population undiscovered. In this paper, we attempt to identify any Galactic sources amongst the \citet{Becker2010} sample by measuring their proper motions as a first step to finding quiescent BHXB candidates. Amongst the 33 targets, we could measure the proper motion of six sources. We find that G32.7193-0.6477 is a Galactic source and are able to constrain the parallax of this source with a 3$\sigma$ significance. We found three strong Galactic candidates, G32.5898-0.4468, G29.1075-0.1546, and G31.1494-0.1727, based purely on their proper motions, and suggest that G29.1075-0.1546, is also likely Galactic. We detected two resolved targets for multiple epochs (G30.1038+0.3984 and G29.7161-0.3178). We find six targets are only detected in one epoch and have an extended structure. We cross-match our VLBA detections with the currently available optical, infrared and X-ray surveys, and did not find any potential matches. We did not detect 19 targets in any VLBA epochs and suggest that this could be due to limited $uv$-coverage, drastic radio variability or faint, extended nature of the sources.
\end{abstract}



\section{Introduction}\label{Section 1}
In the past few years, new techniques and wide field surveys in various parts of the electromagnetic spectrum have been adopted to search for non-interacting black holes and/or quiescent black hole X-ray binaries (BHXBs) in our Galaxy in an effort to put together a complete sample of black holes (BHs). For instance, The Swift Bulge Survey, an X-ray survey of the Galactic bulge \citep{Shaw2020,Bahramian2021}, was designed to detect Very Faint X-ray transients \citep[VFXTs;][]{Wijnands2006}. This is a class of X-ray transients with outbursts that are not brighter than 10$^{36}$\,erg\,s$^{-1}$. There have also been multiple efforts in the field of optical spectroscopy that have involved measuring the radial velocity of the optically bright component in a binary and determining the orbital parameters and ultimately the mass function of the binary. Such studies have led to the discovery of possible triple systems, with a BH candidate as one of its members \citep{Thompson2019,van2020,Thompson2020,Eldridge2019,Rivinius2020,el-badry2020}. Although the nature of these systems is currently under debate \citep{Bodensteiner2020}, these discoveries have opened up the possibility of using spectroscopy as a novel way to detect triple systems where the inner binary is non-accreting binary BH system \citep{Hayashi2020}. A similar photometric and spectroscopic survey of 25 globular clusters with the Multi Unit Spectroscopic Explorer (MUSE) on the European Southern Observatory's Very Large Telescope (VLT) resulted in the detection of three detached binary systems harbouring a star and a stellar mass black hole candidate in one of the globular clusters included in the survey, NGC 3201 \citep{Giesers2018,Giesers2019}. \par
\citet{Casares2018a} proposed the use of photometry to study the H$_{\alpha}$ emission line with a combination of narrow and broad filters to search for quiescent BHXBs in our Galaxy. If the width of the H$_{\alpha}$ emission line, which is a tell-tale sign of an accretion disc, could be determined with high enough accuracy and combined with the orbital period of the binary, it would help in determining the mass of the accreting object. Hence, a Galactic plane survey based on the above technique could yield at least 50 quiescent black holes \citep{Casares2018b}, almost doubling the number of BHXBs that we know of today \citep{Corral-Santana2016,Tetarenko2016a}. \par
Other than optical/IR and X-rays, quiescent BHXBs can also be searched for at radio wavelengths \citep{Maccarone2005,Maccarone2018}. Radio emission in quiescent BHXBs is expected to arise from compact relativistic jets \citep{Blandford1979ApJ} that should be unresolved even with the currently available high resolution radio interferometers. It has been observed that X-ray ($L_{\rm{X}}$) and radio ($L_{\rm{R}}$) luminosities of BHXBs in a quiescent state show a correlation given by $L_{\rm{R}}\propto\,L_{\rm{X}}^{0.61\pm0.03}$ \citep[][]{Gallo2018}, which suggests that the radio flux density of these systems falls off slower than the X-ray flux as the accretion rate drops. Searches for quiescent BHXBs in our Galaxy at radio wavelengths could then be a highly effective method to observe this otherwise elusive population. The Milky-way ATCA and VLA Exploration of Radio-sources in Clusters \citep[MAVERIC; eg.,][]{Shishkovsky2018,Bahramian2018} survey is one such survey, although it aims at finding X-ray binaries in globular clusters, where their formation rate is significantly higher than the field. The work we present in this paper is a systematic attempt to identify quiescent BHXBs in our Galaxy in the radio band.
\subsection{A radio variability survey - \citet{Becker2010}}
A search for variable radio sources in the Galactic plane was conducted by \citet{Becker2010}, B10 hereon, by comparing 6\,cm (5\,GHz) radio flux densities of sources from two surveys using the VLA. The first epoch for this variability study was taken from a survey done between 1989 and 1991 \citep{Becker1994}, and the second and third epochs were data observed under the Co-Ordinated Radio 'N' Infrared Survey for High-mass star formation \citep[CORNISH;][]{Purcell2013} survey in 2005--2006. This allowed them to study the radio variability of sources with a time baseline of 15 years. The first epoch was observed in the C and BnC configurations, whereas the second and the third epochs were in the B configuration. The differing observing configurations and resolutions could result in spurious variability in extended objects. Limiting their study to regions that were covered in at least two out of the three epochs and defining their variability criterion as targets that showed a difference in their peak flux densities greater than five times r.m.s uncertainty at the two epochs, they found 39 variable sources. Comparing the variability of these sources to an extragalactic radio variability study \citep{de2004}, which was done using the 1995 and 2002 epochs of the FIRST survey \citep{Becker1995}, they were able to suggest that 6 sources (out of 39) with the lowest variability could be background radio sources. The remaining 33 highly variable sources were suggested to be a part of a population of variable Galactic radio sources, based on their high concentration towards the Galactic Plane.\par
Although the nature of these 33 sources was unclear, B10 were able to rule out a few scenarios. The non-detection of bright optical counterparts for these targets strongly argues against radio flaring stars. The most radio-luminous dMe flare stars reach 1\,mJy only at distances $<$13 pc. Allowing for flares of a factor of 500 \citep[e.g.][]{Osten2008} only permits detection out to 290 pc; such a population would be associated with bright optical/IR sources, and would not be concentrated in the Galactic Plane and Bulge like the B10 sample. Radio masers were ruled out because no known atomic or molecular transitions occur in the observed bandpass. Pulsars were also ruled out as possible candidates of the B10 sample, since even the weakest of pulsars would have been detected in low-frequency pulsar surveys, given the steep spectrum and the high flux densities of these objects at 1.4 GHz. GRBs and supernovae were excluded due to the quick rise time of their expected radio flux densities (a few days), which would be very short as compared to the 15-year variability timescale observed for the sources in the B10 sample.  \par
B10 could not rule out that a subset of the highly variable B10 sources could be X-ray binaries, possibly black hole X-ray binaries (BHXBs). Most BHXBs were first discovered when they were in an outburst. During this stage, the rate of accretion onto the black hole increases, resulting in brightening across the electromagnetic spectrum. This method of detecting BHXBs creates a bias in the sample of studied systems against the larger number of systems that are accreting at a low rate and exist only in a quiescent state, (L$_{X}<$10$^{33.5}$\,ergs\,s$^{-1}$; \citealp{McClintock2006}). It also creates a bias against ultra compact and short orbital sources \citep{Arur2018} that have peak outburst luminosities lower than the triggering threshold for current all-sky X-ray monitors. Since most BHXBs spend the majority of their lifetime in a quiescent state, dependence on outburst activity for their discovery also limits the rate at which they are being detected \citep{Tetarenko2016a,Corral-Santana2016}. We are also biased against observing the most massive black holes in BHXBs as evidenced by the number of known and dynamically confirmed BHXBs found close to the Galactic Plane and the Galactic Centre \citep{Jonker2021}. The stark difference between the number of BHXBs expected in our Galaxy \citep[$10^4$;][]{Jonker2011}, the number of known BHXB candidates \citep[77;][]{Tetarenko2016a} and the number of BHXBs detected during quiescence in the radio wavelengths \citep[7; e.g.,][]{Plotkin2021,Carotenuto2022} indicates that a large population of BHXBs are likely hidden from conventional discovery techniques. Since none of the sources from the B10 sample have been detected as outbursting X-ray or optical sources, this suggests that some of these could be quiescent BHXBs.
\subsection{Motivation for a Very Long Baseline Array campaign}
\citet{Kirsten2014} made a serendipitous discovery of a foreground radio source whilst studying the proper motions of compact radio sources towards the globular cluster M15. They found that one of the sources in their field of view was in fact a foreground Galactic object based on the proper motion and parallax measurement to the source. This parallax measurement put the source at a distance of 2.2$\pm$0.5\,kpc, which was later updated to 2.4$^{3.4}_{-1.8}$\,kpc \citep{Atri2019}. The distance measurement to the source enabled the distinction of the source from an object in the M15 globular cluster, which is known to be at a distance of 10.3$\pm$0.4\,kpc \citep{Bosch2006}. Follow-up of this source by \citet{Tetarenko2016} suggested the presence of a red stellar counterpart in the mass range of 0.1--0.2\,M$_{\odot}$. They claimed that the binary companion of this red stellar object is possibly a BHXB discovered in quiescence, which given the area of the sky surveyed to make the original detection by \citet{Kirsten2014}, implies that there should be around 10$^4$--10$^8$ such systems in our Galaxy. This was an increase in the number of expected BHXBs in our Galaxy from what was suggested by population synthesis models and surveys \citep[10$^2$--10$^4$;][]{Romani1998,Pfahl2003,Corral-Santana2016}. Thus, this serendipitous discovery of a radio source and candidate quiescent BHXB in the foreground of the globular cluster M15 suggests that there could be more such systems hiding in the Galaxy. \par
This was our motivation for conducting a systematic search for Galactic systems in the B10 sample of 33 highly-variable sources. In this paper, we detail the feasibility of using high resolution radio interferometry to distinguish between Galactic and extragalactic radio sources from the sample of highly variable radio sources in the B10 study, and potentially find new quiescent BHXBs in the sample of Galactic sources. Galactic sources, except for the small population that only move radially compared to Earth, will have non-zero proper motions, whereas an extragalactic background source will be stationary with respect to the International Celestial Reference Frame \citep[ICRF;][]{Ma1998}, at least to the capability of current telescopes. Very Long Baseline Interferometry (VLBI) allows us to measure the proper motion, and in some cases the parallax, of compact radio sources if their position is measured for two or more epochs separated by several months. Most neutron stars and BHXBs receive natal kicks when the compact object in the binary is formed, giving them a higher velocity than the local standard of rest \citep[e.g.][]{Blaauw1961,Chugai1984,Janka2013}. This kick might manifest itself in the form of proper motions larger than the motion due to only the Galactic rotation, making high proper motion Galactic radio sources more likely to be X-ray binaries \citep[e.g.,][]{Dhawan2007,Mirabel2001} or pulsars \citep{Verbunt2017}. \par
In this paper, we report our findings from a proper motion measurement campaign using the Very Long Baseline Array (VLBA) to target the sample of the highly variable B10 sources. We describe our observation setup in Section \ref{Section 2}, followed by a description of the steps we followed to reduce the data in Section \ref{Section 3}. In Section \ref{Section 4} we provide the results of the data reduction ,and report the positions and flux density measurements of the detections. We discuss our analysis to measure the proper motions and parallaxes of our targets in Section \ref{Section 5}, and the implications of these measurements in Section \ref{Section 6}.   

\section{Observations}\label{Section 2}
B10 suggested that 33 sources (out of 39 targets in their sample) with a fractional variability (which they defined as $\mathit{f}$\,$\equiv$\,Ratio of highest observed peak flux density to the lowest observed peak flux density) of $>$2 could be Galactic sources. We chose to observe these 33 targets for our VLBA campaign having discarded the six targets with $f\leq$1.5 as potential extragalactic radio sources. Our observations were part of a VLBA filler program, program code BH208, which was supposed to observe the 33 targets for three epochs with the epochs spread over a $\approx$1-year period. As this was a filler program, depending upon the availability of the telescope, some sources were only observed for two epochs and only had a couple of months' separation between the two epochs. Two epochs are still enough to constrain proper motion, although it is difficult to estimate the level of contribution in this motion due to the unknown parallax of the source. Nonetheless, any significant motion, whether due to parallax or proper motion, would imply a Galactic nature.\par
The observations were planned such that two sources could be observed in a one hour observing block, with 10-15 minutes of total on-source time split in two blocks to maximise $uv$-coverage, and a total of 6-10 minutes on the phase reference calibrator for each target. A standard phase referencing scheme was followed to obtain sub-milliarcsecond (mas) level precision on the absolute positions of the targets. The observing blocks were scheduled to switch between target (two minutes) and phase calibrator (one minute) for a cycle time of three minutes, and bracketed by a few minutes on a fringe finder at the start and end of the observing block. The observation IDs start from BH208A and run through to BH208O in order of decreasing Galactic longitude (see Table \ref{Observations_VLBA_survey}). All observations were carried out in the X band with a central frequency of 8.4\,GHz, to maximise sensitivity and astrometric precision, while not typically being strongly affected by scattering in the Galactic plane. We used a bandwidth of 256 MHz per polarisation and a data recording rate of 2\,Gbps for all observations. Depending on the target-calibrator separation the observations took three different forms. Most of the observing blocks had a phase reference calibrator that was shared between the two nearby targets. A few observation blocks had separate calibrators for each target, i.e., two calibrators and two targets (Obs ID - BH208C and BH208I). Three observing blocks (Obs ID- BH208A, BH208D and BH208L) had three targets each, wherein two of the targets (separation of $<$1$^{\prime\prime}$) fell in the same VLBA primary beam. In such cases we observed one phase calibrator for the two targets in the same VLBA beam. We also used the multi-phase centre correlation capabilities of DiFX for such pairs of targets i.e., we used the mid-point between these two targets as the pointing centre and then requested correlation of the data at the positions of both the targets. The third target had its own phase reference calibrator. All ten VLBA antennas were used for the observations, unless impacted by weather or other technical problems. All the data were correlated using the DiFX software correlator \citep{Deller2011}. \par

\begin{table}
  \centering
  \caption{Summary of the filler VLBA observing campaign. The Obs ID column has BH208 preceding all the reported Obs IDs (e.g., BH208A). The target names are derived from Galactic coordinates (also used by B10 survey) and the names in parentheses are the shortened names that we adopt for the rest of this paper. The name is shortened by combining the observing ID and the number of the target in the observing block. The date of observation is given as dd/mm/yy, where yy is the number of year after the year 2000.}

    \begin{tabular}{llllll}
    Obs ID & Target name &  \multicolumn{3}{c}{Dates of observations} \\
    (BH208)      &       & \multicolumn{1}{l}{Epoch 1} & \multicolumn{1}{l}{Epoch 2} & \multicolumn{1}{c}{Epoch 3} \\ \hline
    A     & G39.1105-0.0160 (A\romannumeral 1) & 14/08/15 & 26/12/15 & \multicolumn{1}{l}{-} \\
    A     & G37.7347-0.1126 (A\romannumeral 2) & 14/08/15 & 26/12/15 & \multicolumn{1}{l}{-} \\
    A     & G37.7596-0.1001 (A\romannumeral 3) & 14/08/15 & 26/12/15 & \multicolumn{1}{l}{-} \\
    B     & G37.2324-0.0356 (B\romannumeral 1) & 31/05/15 & 09/11/15 & 31/01/16 \\
    B     & G32.7193-0.6477 (B\romannumeral 2) & 31/05/15 & 09/11/15 & 31/01/16 \\
    C     & G32.5898-0.4468 (C\romannumeral 1) & 31/05/15 & 02/01/16 & \multicolumn{1}{l}{-} \\
    C     & G31.1595+0.0449 (C\romannumeral 2) & 31/05/15 & 02/01/16 & \multicolumn{1}{l}{-} \\
    D     & G31.1494-0.1727 (D\romannumeral 1) & 15/08/15 & 30/12/15 & \multicolumn{1}{l}{-} \\
    D     & G30.4376-0.2062 (D\romannumeral 2) & 15/08/15 & 30/12/15 & \multicolumn{1}{l}{-} \\
    D     & G30.4460-0.2148 (D\romannumeral 3) & 15/08/15 & 30/12/15 & \multicolumn{1}{l}{-} \\
    E     & G30.1038+0.3984 (E\romannumeral 1) & 01/06/15 & 06/12/15 & 04/02/16 \\
    E     & G29.7161-0.3178 (E\romannumeral 2) & 01/06/15 & 06/12/15 & 04/02/16 \\
    F     & G29.5779-0.2685 (F\romannumeral 1) & 06/09/15 & 09/01/16 & 31/01/16 \\
    F     & G29.4959-0.3000 (F\romannumeral 2) & 06/09/15 & 09/01/16 & 31/01/16 \\
    G     & G29.6051-0.8590 (G\romannumeral 1) & 27/09/15 & 11/01/15 & \multicolumn{1}{l}{-} \\
    G     & G29.1978-0.1268 (G\romannumeral 2) & 27/09/15 & 11/01/15 & \multicolumn{1}{l}{-} \\
    H     & G29.1075-0.1546 (H\romannumeral 1) & 24/10/15 & 15/01/16 & \multicolumn{1}{l}{-} \\
    H     & G28.6204-0.3436 (H\romannumeral 2) & 24/10/15 & 15/01/16 & \multicolumn{1}{l}{-} \\
    I     & G27.8821+0.1834 (I\romannumeral 1) & 12/07/15 & 04/01/16 &  \multicolumn{1}{l}{-}\\
    I     & G26.2818+0.2312 (I\romannumeral 2) & 12/07/15 & 04/01/16 &  \multicolumn{1}{l}{-}\\
    J     & G26.0526-0.2426 (J\romannumeral 1) & 31/07/15 & 27/12/15 & \multicolumn{1}{l}{-} \\
    J     & G25.7156+0.0488 (J\romannumeral 2) & 31/07/15 & 27/12/15 & \multicolumn{1}{l}{-} \\
    K     & G25.4920-0.3476 (K\romannumeral 1) & 19/07/15 & 24/12/15 & \multicolumn{1}{l}{-}  \\
    K     & G25.2048+0.1251 (K\romannumeral 2) & 19/07/15 & 24/12/15 & \multicolumn{1}{l}{-}  \\
    L     & G24.5343-0.1020 (L\romannumeral 1) & 01/11/15 & 20/01/16 &  \multicolumn{1}{l}{-} \\
    L     & G24.5405-0.1377 (L\romannumeral 2) & 01/11/15 & 20/01/16 & \multicolumn{1}{l}{-}\\
    L     & G24.3367-0.1574 (L\romannumeral 3) & 01/11/15 & 20/01/16 & \multicolumn{1}{l}{-}\\
    M     & G23.6644-0.0372 (M\romannumeral 1) & 10/08/15 & 02/01/16 &\multicolumn{1}{l}{-} \\
    M     & G23.4186+0.0090 (M\romannumeral 2) & 10/08/15 & 02/01/16 & \multicolumn{1}{l}{-} \\
    N     & G23.5585-0.3241 (N\romannumeral 1) & 22/07/15 & 24/12/15 & \multicolumn{1}{l}{-} \\
    N     & G22.9743-0.3920 (N\romannumeral 2) & 22/07/15 & 24/12/15 & \multicolumn{1}{l}{-} \\
    O     & G22.9116-0.2878 (O\romannumeral 1) & 01/06/15 & 13/12/15 & \multicolumn{1}{l}{-} \\
    O     & G22.7194-0.1939 (O\romannumeral 2) & 01/06/15 & 13/12/15 &  \multicolumn{1}{l}{-} \\
    \hline
    \end{tabular}%
    \label{Observations_VLBA_survey}
\end{table}%

\section{Data reduction} \label{Section 3}
The data were reduced using the Astronomical Image Processing System \citep[AIPS 31DEC17;][]{Greisen2003} and some imaging steps were carried out using {\tt{DIFMAP}} \citep{Shepherd1994}. This VLBA survey includes 72 datasets, and to minimise the time needed to process the datasets the data were reduced in the following two stages.
\subsection{First order reduction}
The first-order reduction included obtaining the images of each target using a pipeline and determining whether the target was detected in those images. A custom pipeline was written in ParselTongue\footnote{A \textsc{python} wrapper to AIPS} \citep{Kettenis2006} to go through the data reduction steps in AIPS starting with loading the data into AIPS, followed by correcting for updated Earth Orientation Parameters and correcting for the ionospheric and dispersive delays using a -riori measurements of total electron content. We then used a bright fringe finder to correct for the instrumental phases and delays, and for the bandpass calibration. A final amplitude correction was done by using the a-priori system temperatures before splitting off the phase calibrator. A model of the phase calibrator was made by iteratively self-calibrating the phases and reducing the solution interval in each iteration, until a minimum solution interval of 30 seconds was reached. Amplitude self-calibration was also done using a solution interval of 10 minutes. This gave a best model of the calibrator, which we then used to run {\tt{FRING}} again and determine the delay and rate calibration to be applied to the target. The calibrated target data were finally split from the main dataset to image individually. Since natural weighting is most sensitive but has lower natural image resolution and stronger sidelobes, we tested different robustness factors from 0 (intermediate between natural and uniform weighting) to 5 (natural weighting) \citep{Briggs1995}. If a target was detected (a bright fringe or a $\>5\sigma$ detection during imaging, see Section \ref{subsec:Imaging}) in one or more epochs, then we retained that source to perform a more careful calibration.
\subsection{Second order reduction}\label{second_order_red}
The sources that were detected with a $\geq$5$\sigma$ significance in one or more epochs were chosen to go through a second-order data reduction process. A pipeline was written to stack the phase calibrator data for different epochs to improve the calibrator model and to make sure that, regardless of $uv$-coverage, the same calibrator model was used to determine the delay and rate solutions to be interpolated to the target in all epochs. The best possible model of this stacked calibrator was made by iteratively using self-calibration and reducing the self-calibration solution intervals with each iteration, until we reached a minimum solution interval of 30 seconds. After the phase self-calibration, a single iteration of amplitude self-calibration was done using a solution interval of 10 minutes. This model was then used as an input to {\tt{FRING}} that gave the delay and rate solutions that were then applied to the target epoch-wise. All data below 15$^{\circ}$ elevation were flagged to reduce systematic errors due to low elevations. Usually, a check source, which is a compact extragalactic radio source, is observed to gauge the effect of systematics on the quality of the phase solutions applied to the target. Since these short, 1-hour observations were part of a filler program, we did not have time to observe a check source. In such cases, a few scans of the phase calibrator can be treated as a second target source and the change in the position of the calibrator between these scans and those that were self-calibrated can be measured. We used every second phase calibrator scan to be imaged as a check source, after applying the phase and delay solutions to these scans as derived from the remaining phase calibrator scans. The variation in the astrometry of the check source between epochs is used to estimate the extent of systematic errors on the position measurement of the target (see Section \ref{Section 5}).
\begin{figure*}
\begin{center}
\begin{tabular}{cc}
\phantom{123456} A\romannumeral 3 & \phantom{123456} F\romannumeral 2 \\
\includegraphics[width=8cm,height=7cm,angle=0]{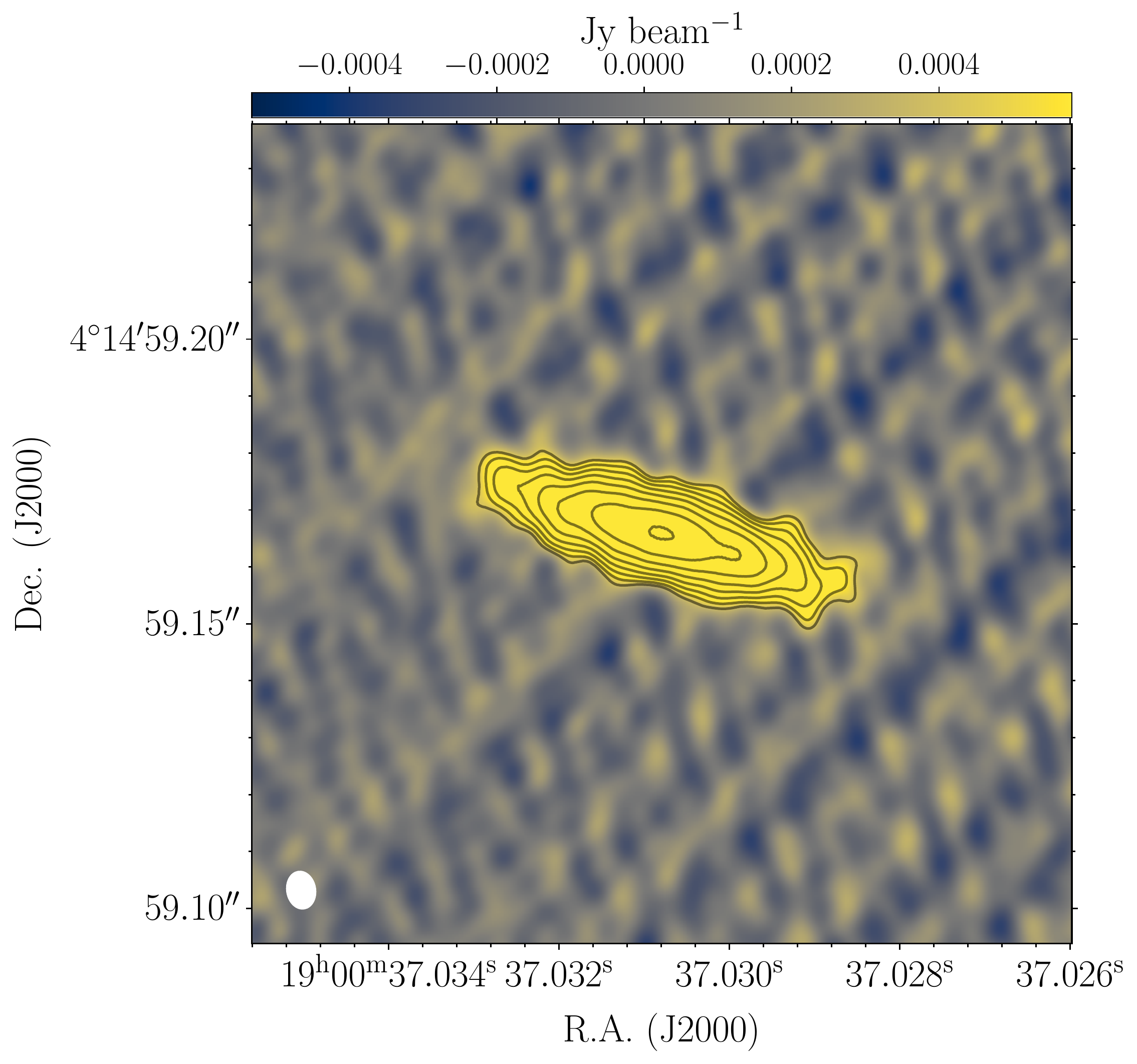} &
\includegraphics[width=8cm,height=7cm,angle=0]{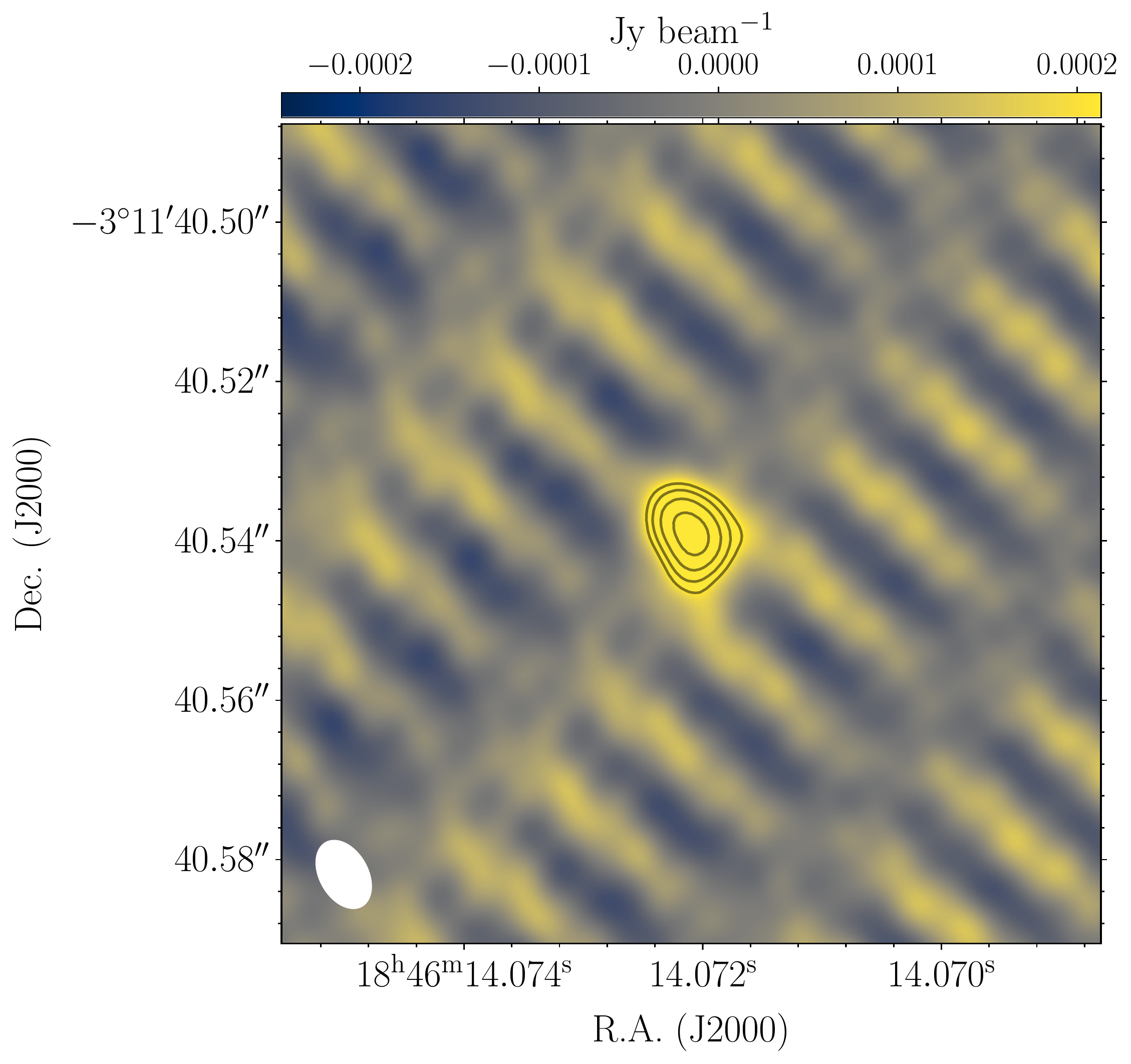}  \\
\end{tabular}
\end{center}
\caption{ \label{A3_F2_contour} Image and contours of the Gaussian elliptical model fit to the two single-epoch detections of targets G37.7596-0.1001 (A\romannumeral 3; \textit{left panel}) and G29.4959-0.3000 (F\romannumeral 2; \textit{right panel}). The white ellipses in the left-bottom corner of each image represents the beam size and shape. The contours levels are in steps of rms of each image with increments of sqrt(2), with the lowest contour at 3$\times$rms.  }
\end{figure*}
\begin{table*}
  \centering
  \caption{Detection summary of the VLBA observations conducted at 8.4\,GHz along with the VLA detections done at 5\,GHz reported in B10. We report the peak intensity, $S_{pk}$, and the integral intensity ,$S_{int}$, for the resolved sources, and only the $S_{pk}$ for the compact sources. The rms is reported with the peak intensity, and the uncertainty reported with $S_{int}$ is the uncertainty in the fit. We use a superscript `R' to note that the target is resolved and has been imaged after $uv$-tapering. Some targets were also imaged after concatenating datasets from two epochs, these targets can be identified by their `D' superscript. As the detections of these targets are unreliable in the second epoch and the data have been concatenated with the first epoch we have indicated that with a `+Epoch 1' in the table. `Invalid data' are the observations that did not have the correct correlation position and thus could not be used in the analysis; `Fr' are targets where there was a strong fringe pattern in the image but we could not ascertain exactly where in the fringe the source was located. We report the integral intensity of the VLA detections of these sources as reported in B10. Epoch 1 are observations taken in 1990s, whereas Epoch 2 and 3 were taken in the years 2005 and 2006. }
    \begin{tabular}{lccccccccl}
     Target name & \multicolumn{6}{c}{VLBA detections (at 8.4\,GHz)} & \multicolumn{3}{c}{VLA detections (at 5\,GHz)} \\
                 & \multicolumn{2}{c}{Epoch 1} & \multicolumn{2}{c}{Epoch 2} & \multicolumn{2}{c}{Epoch 3}  & \multicolumn{1}{c}{Epoch 1} & \multicolumn{1}{c}{Epoch 2} & \multicolumn{1}{c}{Epoch 3}\\
                 & $S_{pk}$  &    $S_{int}$ & $S_{pk}$ &    $S_{int}$       & $S_{pk}$  &    $S_{int}$       &    $S_{int}$                &    $S_{int}$        &    $S_{int}$   \\
                 & (rms)  &    (uncertainty)       & (rms)  &     (uncertainty)       &  (rms)  &     (uncertainty)        &    (uncertainty)                & (uncertainty)  &    (uncertainty)       \\
                 & $\upmu$Jy\,bm$^{-1}$  &    $\upmu$Jy       & $\upmu$Jy\,bm$^{-1}$  &    $\upmu$Jy       &  $\upmu$Jy\,bm$^{-1}$  &    $\upmu$Jy        &    $\upmu$Jy\,bm$^{-1}$                  & $\upmu$Jy   &    $\upmu$Jy\,bm$^{-1}$       \\
                 \hline
     A\romannumeral 1 & \multicolumn{1}{l}{624\phantom{1} (95)} &\multicolumn{1}{c}{-} & \multicolumn{1}{l}{779\phantom{1} (126)}   & \multicolumn{1}{c}{-} & \multicolumn{1}{c}{-}     & \multicolumn{1}{c}{-} & \multicolumn{1}{l}{900\phantom{111}(190)}  & \multicolumn{1}{c}{-}                 & \multicolumn{1}{l}{4200\phantom{1} (320)} \\
     A\romannumeral 2 & \multicolumn{1}{l}{\phantom{1234} (122)} &\multicolumn{1}{c}{-} & \multicolumn{1}{l}{Invalid data}   &\multicolumn{1}{l}{Invalid data}  & \multicolumn{1}{c}{-}      &\multicolumn{1}{c}{-}     & \multicolumn{1}{l}{$<$1300 (630)}         & \multicolumn{1}{c}{-}                 & \multicolumn{1}{l}{12200 (320)}  \\
     A\romannumeral 3$^{\mathrm{R}}$ & \multicolumn{1}{l}{4720 (142)}   & \multicolumn{1}{c}{53613 (1750)} & \multicolumn{1}{l}{Invalid data}  & \multicolumn{1}{l}{Invalid data}  & \multicolumn{1}{c}{-}   & \multicolumn{1}{c}{-}       & \multicolumn{1}{l}{$<$1500 (740)}         & \multicolumn{1}{c}{-}                 & \multicolumn{1}{l}{12000 (320)}   \\
     B\romannumeral 1 & \multicolumn{1}{l}{\phantom{1234} (91)}      & \multicolumn{1}{c}{-}   & \multicolumn{1}{l}{\phantom{1234} (73)}  & \multicolumn{1}{c}{-}    & \multicolumn{1}{c}{\phantom{1234} (82)}       & \multicolumn{1}{c}{-}   & \multicolumn{1}{l}{2400\phantom{12}(270)}            & \multicolumn{1}{c}{-}                 & \multicolumn{1}{l}{$<$500\phantom{1}(270)} \\
     B\romannumeral 2 & \multicolumn{1}{l}{563\phantom{1} (99)}      & \multicolumn{1}{c}{-}   & \multicolumn{1}{l}{599\phantom{1} (94)}  & \multicolumn{1}{c}{-}    & \multicolumn{1}{l}{729\phantom{1} (100)}  & \multicolumn{1}{c}{-}   & \multicolumn{1}{c}{-}                     & \multicolumn{1}{l}{4300\phantom{1} (170)}        & \multicolumn{1}{l}{2000\phantom{1} (320)}\\
     C\romannumeral 1 & \multicolumn{1}{l}{1032 (131)}               & \multicolumn{1}{c}{-}   & \multicolumn{1}{l}{1075 (114)}           & \multicolumn{1}{c}{-}    & \multicolumn{1}{c}{-}          & \multicolumn{1}{c}{-}   & \multicolumn{1}{c}{-}                     & \multicolumn{1}{l}{3600\phantom{1} (170)}        & \multicolumn{1}{l}{$<$700\phantom{1}(330)}\\
     C\romannumeral 2 & \multicolumn{1}{l}{\phantom{1234} (71)}      & \multicolumn{1}{c}{-}   & \multicolumn{1}{l}{\phantom{1234} (63)}  & \multicolumn{1}{c}{-}    & \multicolumn{1}{c}{-}          & \multicolumn{1}{c}{-}   & \multicolumn{1}{l}{11600\phantom{1}(320)}           & \multicolumn{1}{l}{18800 (230)}       & \multicolumn{1}{l}{17200 (350)}\\
     D\romannumeral 1 & \multicolumn{1}{l}{766\phantom{1} (148)}     & \multicolumn{1}{c}{-}   & \multicolumn{1}{l}{761\phantom{1} (141)} & \multicolumn{1}{c}{-}    & \multicolumn{1}{c}{-}          & \multicolumn{1}{c}{-}   & \multicolumn{1}{l}{1600\phantom{12}(260)}            & \multicolumn{1}{l}{5600\phantom{1} (190)}        & \multicolumn{1}{l}{1800\phantom{1} (350)}\\
     D\romannumeral 2 & \multicolumn{1}{l}{\phantom{1234} (106)} & \multicolumn{1}{c}{-}   & \multicolumn{1}{l}{Invalid data}         & \multicolumn{1}{l}{Invalid data}    & \multicolumn{1}{c}{-}          & \multicolumn{1}{c}{-}   & \multicolumn{1}{l}{1200\phantom{12}(270)}            & \multicolumn{1}{l}{13100 (210)}       & \multicolumn{1}{l}{17500 (320)} \\
     D\romannumeral 3 & \multicolumn{1}{l}{\phantom{1234} (130)}                        & \multicolumn{1}{c}{-}   & \multicolumn{1}{l}{Invalid data}         & \multicolumn{1}{l}{Invalid data}    & \multicolumn{1}{c}{-}          & \multicolumn{1}{c}{-}   & \multicolumn{1}{l}{3600\phantom{12}(320)}            & \multicolumn{1}{l}{6600\phantom{1} (220)}        & \multicolumn{1}{l}{6600\phantom{1} (310)}\\
     E\romannumeral 1 & \multicolumn{1}{l}{967\phantom{1} (65)}  &  \multicolumn{1}{l}{1562\phantom{1} (157)}   & \multicolumn{1}{l}{939\phantom{1} (56)}  & \multicolumn{1}{l}{1830 (159)}  & \multicolumn{1}{l}{902\phantom{1} (51)} & \multicolumn{1}{l}{1793 (145)} & \multicolumn{1}{l}{3700\phantom{12}(460)}            & \multicolumn{1}{l}{6800\phantom{1} (200)}        & \multicolumn{1}{l}{7700\phantom{1} (320)}\\
     E\romannumeral 2 & \multicolumn{1}{l}{2150 (71)}  &  \multicolumn{1}{l}{9693\phantom{1} (383)} & \multicolumn{1}{l}{2328 (106)}   &\multicolumn{1}{l}{9286 (502)}& \multicolumn{1}{l}{2330 (90)} &\multicolumn{1}{l}{9107 (430)} & \multicolumn{1}{l}{16600\phantom{1}(260)}           & \multicolumn{1}{l}{30000 (240)}       & \multicolumn{1}{l}{30100 (370)}\\
     F\romannumeral 1 & \multicolumn{1}{c}{\phantom{123}(61)} & \multicolumn{1}{c}{-}  & \multicolumn{1}{c}{\phantom{1234} (169)} & \multicolumn{1}{c}{-} & \multicolumn{1}{c}{\phantom{123} (71)} & \multicolumn{1}{c}{-} & \multicolumn{1}{l}{7600\phantom{12}(400)}            & \multicolumn{1}{l}{11500 (240)}       & \multicolumn{1}{l}{5100\phantom{1} (340)} \\
     F\romannumeral 2$^{\mathrm{R}}$ & \multicolumn{1}{l}{763\phantom{1} (74)} & \multicolumn{1}{l}{1156\phantom{1} (172)} &\multicolumn{1}{c}{\phantom{1234} (147)} & \multicolumn{1}{c}{-} & \multicolumn{1}{c}{\phantom{123} (75)} & \multicolumn{1}{c}{-} &\multicolumn{1}{l}{5200\phantom{12}(160)}&\multicolumn{1}{l}{13500 (200)}    & \multicolumn{1}{l}{7500\phantom{1} (320)} \\
     G\romannumeral 1 & \multicolumn{1}{l}{\phantom{1234} (65)} & \multicolumn{1}{c}{-} & \multicolumn{1}{c}{\phantom{123} (78)}  & \multicolumn{1}{c}{-} & \multicolumn{1}{c}{-} & \multicolumn{1}{c}{-} & \multicolumn{1}{c}{-}    & \multicolumn{1}{l}{2500\phantom{1} (230)}        & \multicolumn{1}{l}{8300\phantom{1} (340)}\\
     G\romannumeral 2$^{\mathrm{R,D}}$ & \multicolumn{1}{l}{274\phantom{1} (45)} & \multicolumn{1}{l}{536\phantom{12} (126)} & \multicolumn{1}{l}{+Epoch 1}   & \multicolumn{1}{l}{+Epoch 1} & \multicolumn{1}{c}{-}    & \multicolumn{1}{c}{-} & \multicolumn{1}{l}{2000\phantom{12}(250)}            & \multicolumn{1}{l}{4100\phantom{1} (220)}        & \multicolumn{1}{l}{1700\phantom{1} (340)}\\
     H\romannumeral 1 & \multicolumn{1}{l}{912\phantom{1} (142)}  & \multicolumn{1}{c}{-}  & \multicolumn{1}{l}{844\phantom{1} (120)}   & \multicolumn{1}{c}{-}  & \multicolumn{1}{c}{-} & \multicolumn{1}{c}{-}  & \multicolumn{1}{l}{6500\phantom{12}(130)} & \multicolumn{1}{l}{11400 (230)}       & \multicolumn{1}{l}{10500 (330)}\\
     H\romannumeral 2 & \multicolumn{1}{l}{1101 (127)} & \multicolumn{1}{c}{-}   & \multicolumn{1}{l}{832\phantom{1} (109)} & \multicolumn{1}{c}{-}  & \multicolumn{1}{c}{-} & \multicolumn{1}{c}{-} & \multicolumn{1}{l}{3300\phantom{12}(180)}            & \multicolumn{1}{l}{6400\phantom{1} (220)}        & \multicolumn{1}{l}{8400\phantom{1} (310)}\\
     I\romannumeral 1$^{\mathrm{R,D}}$ & \multicolumn{1}{l}{903\phantom{1} (50)} & \multicolumn{1}{l}{2157\phantom{1} (162)} & \multicolumn{1}{l}{+Epoch 1}  & \multicolumn{1}{l}{+Epoch 1} & \multicolumn{1}{c}{-} & \multicolumn{1}{c}{-}  & \multicolumn{1}{l}{3300\phantom{12}(210)}            & \multicolumn{1}{c}{-}                 & \multicolumn{1}{l}{5800\phantom{1} (320)}\\
     I\romannumeral 2$^{\mathrm{R,D}}$ & \multicolumn{1}{l}{1129 (59)} & \multicolumn{1}{l}{4542\phantom{1} (290)} & \multicolumn{1}{l}{+Epoch 1} & \multicolumn{1}{l}{+Epoch 1} & \multicolumn{1}{c}{-} & \multicolumn{1}{c}{-} & \multicolumn{1}{l}{9000\phantom{12}(270)}            & \multicolumn{1}{c}{-}                 & \multicolumn{1}{l}{14300 (290)}\\
     J\romannumeral 1 & \multicolumn{1}{l}{Fr\phantom{12} (186)} & \multicolumn{1}{c}{-}  & \multicolumn{1}{l}{Fr\phantom{12} (130)} & \multicolumn{1}{c}{-} & \multicolumn{1}{c}{-} & \multicolumn{1}{c}{-}    & \multicolumn{1}{l}{7400\phantom{12}(330)}            & \multicolumn{1}{c}{-}                 & \multicolumn{1}{l}{12600 (300)} \\
     J\romannumeral 2 & \multicolumn{1}{l}{\phantom{1234} (80)} & \multicolumn{1}{c}{-}  & \multicolumn{1}{c}{\phantom{12} (88)} & \multicolumn{1}{c}{-}    & \multicolumn{1}{c}{-} & \multicolumn{1}{c}{-}     & \multicolumn{1}{l}{2600\phantom{12}(380)}            & \multicolumn{1}{c}{-}                 & \multicolumn{1}{l}{12600 (370)}\\
     K\romannumeral 1 & \multicolumn{1}{l}{\phantom{1234} (150)}    & \multicolumn{1}{c}{-} & \multicolumn{1}{c}{\phantom{123} (171)} & \multicolumn{1}{c}{-}  & \multicolumn{1}{c}{-} & \multicolumn{1}{c}{-}   & \multicolumn{1}{l}{4300\phantom{12}(170)}            & \multicolumn{1}{c}{-}                 & \multicolumn{1}{l}{1500\phantom{1} (350)}   \\
     K\romannumeral 2 & \multicolumn{1}{l}{\phantom{1234} (135)} & \multicolumn{1}{c}{-} & \multicolumn{1}{l}{\phantom{1234} (142)} & \multicolumn{1}{c}{-} & \multicolumn{1}{c}{-}  & \multicolumn{1}{c}{-}   & \multicolumn{1}{l}{6600\phantom{12}(220)}            & \multicolumn{1}{c}{-}                 & \multicolumn{1}{l}{4200\phantom{1} (320)} \\
     L\romannumeral 1 & \multicolumn{1}{c}{\phantom{1234} (148)} & \multicolumn{1}{c}{-} & \multicolumn{1}{l}{Invalid data}  & \multicolumn{1}{l}{Invalid data} & \multicolumn{1}{c}{-} & \multicolumn{1}{c}{-} & \multicolumn{1}{l}{2800\phantom{12}(520)}            & \multicolumn{1}{c}{-}                 & \multicolumn{1}{l}{4500\phantom{1} (320)} \\
     L\romannumeral 2 & \multicolumn{1}{l}{\phantom{1234} (131)} & \multicolumn{1}{c}{-} & \multicolumn{1}{l}{Invalid data} & \multicolumn{1}{l}{Invalid data} & \multicolumn{1}{c}{-} & \multicolumn{1}{c}{-}  & \multicolumn{1}{l}{1500\phantom{12}(360)}            & \multicolumn{1}{c}{-}                 & \multicolumn{1}{l}{4500\phantom{1} (330)}  \\
     L\romannumeral 3 & \multicolumn{1}{c}{\phantom{1234} (145)} & \multicolumn{1}{c}{-} & \multicolumn{1}{c}{\phantom{1234} (143)}     & \multicolumn{1}{c}{-}  & \multicolumn{1}{c}{-} & \multicolumn{1}{c}{-}  & \multicolumn{1}{l}{2200\phantom{12}(250)}            & \multicolumn{1}{c}{-}                 & \multicolumn{1}{l}{6800\phantom{1} (330)}  \\
     M\romannumeral 1$^{\mathrm{R,D}}$ & \multicolumn{1}{l}{2891 (138)}  & \multicolumn{1}{l}{16464 (916)} & \multicolumn{1}{l}{+Epoch 1} & \multicolumn{1}{l}{+Epoch 1}  & \multicolumn{1}{c}{-}  & \multicolumn{1}{c}{-}  & \multicolumn{1}{l}{5400\phantom{12}(180)}            & \multicolumn{1}{c}{-}                 & \multicolumn{1}{l}{26000 (330)} \\
     M\romannumeral 2 & \multicolumn{1}{c}{\phantom{1234} (92)} & \multicolumn{1}{c}{-}  & \multicolumn{1}{c}{\phantom{123} (81)}  & \multicolumn{1}{c}{-}  & \multicolumn{1}{c}{-} & \multicolumn{1}{c}{-}   & \multicolumn{1}{l}{1800\phantom{12}(400)}            & \multicolumn{1}{c}{-}                 & \multicolumn{1}{l}{5600\phantom{1} (320)} \\
     N\romannumeral 1 & \multicolumn{1}{c}{\phantom{1234} (60)} & \multicolumn{1}{c}{-}  & \multicolumn{1}{c}{\phantom{123} (81)}    & \multicolumn{1}{c}{-}   & \multicolumn{1}{c}{-}   & \multicolumn{1}{c}{-}        & \multicolumn{1}{l}{$<$400\phantom{1}(210)}          & \multicolumn{1}{c}{-}                 & \multicolumn{1}{l}{5500\phantom{1} (300)} \\
     N\romannumeral 2 & \multicolumn{1}{c}{\phantom{1234} (63)} & \multicolumn{1}{c}{-}  & \multicolumn{1}{c}{\phantom{1234} (108)} & \multicolumn{1}{c}{-}   & \multicolumn{1}{c}{-}   & \multicolumn{1}{c}{-}   & \multicolumn{1}{l}{800\phantom{123}(360)}             & \multicolumn{1}{c}{-}                 & \multicolumn{1}{l}{5400\phantom{1} (300)}\\
     O\romannumeral 1 & \multicolumn{1}{c}{\phantom{1234} (81)} & \multicolumn{1}{c}{-}  & \multicolumn{1}{c}{\phantom{123} (53)} & \multicolumn{1}{c}{-}  & \multicolumn{1}{c}{\phantom{1234} (71)}  & \multicolumn{1}{c}{-}   & \multicolumn{1}{l}{3700\phantom{12}(400)}            & \multicolumn{1}{c}{-}                 & \multicolumn{1}{l}{16500 (320)}\\
     O\romannumeral 2 & \multicolumn{1}{c}{\phantom{1234} (73)} & \multicolumn{1}{c}{-}  & \multicolumn{1}{c}{\phantom{123} (61)}& \multicolumn{1}{c}{-}  & \multicolumn{1}{c}{\phantom{1234} (76)} & \multicolumn{1}{c}{-}  & \multicolumn{1}{l}{800\phantom{123}(140)}             & \multicolumn{1}{c}{-}                 & \multicolumn{1}{l}{4100\phantom{1} (320)}\\
     \hline
    \end{tabular}%
\label{Detections_VLBA_survey}
\end{table*}%
\subsection{Imaging}\label{subsec:Imaging}
The single source data files that were split off after applying the calibration steps in Section \ref{second_order_red} were then individually imaged. Images (with natural weighting for the best possible sensitivity) of size 16384 $\times$ 16384 pixels (pixel size of 0.1\,mas) were made to render a large enough image to cater for the large error bars ($\approx$1$^{\prime\prime}$) in the positions of the targets, as detected by the FIRST survey using the VLA \citep{Becker1995}. We made the images with the native 0.5\,MHz frequency resolution without averaging the channels. The edges of the $\approx$1$^{\prime\prime}$ images are affected by 0.06$\%$ bandwidth smearing and 0.17$\%$ time averaging losses, which are negligible. Thus time and frequency smearing did not effect the quality of these $\approx$1$^{\prime\prime}$ images. After determining the location of the bright fringe pattern in the image, the target was imaged again using a smaller image size (8192 $\times$ 8192 pixels) focused on the bright fringe pattern in the larger image. To make sure that the targets detected are compact and are not being resolved out on longer baselines, we imaged all the targets by tapering the long baseline data. {\tt{Difmap}} allows us to use a Gaussian taper (task {\tt{uvtaper}}), which is defined by the taper factor (between 0 and 1) at a specified baseline length. For the purpose of our work, we use a Gaussian taper of 0.5 at a Gaussian radius of 20\,M$\lambda$. That is, we defined a Gaussian profile with a maximum of one (for baselines $\approx$ zero), and a half-width at half-maximum at baselines of 20\,M$\lambda$. A limit of 20\,M$\lambda$ was used as the inner core of VLBA dishes including Kitt Peak, Los Alamos, Owens Valley, Pie Town and Fort Davis have a maximum baseline length of $\approx$20\,M$\lambda$. We used natural weighting (robustness parameter 5) for targets with marginal signal-to-noise ratio (SNR) or the sources that appear resolved, and used robustness parameter 0 where the target had a high SNR and appeared compact. \par
The detection threshold for a source was calculated based on the number of pixels that were searched ($n$), and is given by $(\sqrt{2\ln{n}}\pm0.77/\sqrt{\ln{n}})\times\sigma$ \citep[][ eq. 9.72 and 9.73]{Thompson2017}. Based on this relation and using $n$=8192, we apply a threshold of $6.0\sigma$ to confidently classify the detection as a true source. For detections with a significance of 5--6$\sigma$, we conducted three tests in order to determine whether these detections were astrophysical in nature. 
\begin{itemize}
    \item Test 1: Measured the flux density of the brightest false positive (second brightest pixel) in the dirty image and set the detection threshold 1$\sigma$ above this false positive.
    \item Test 2: Measured the most negative pixel in the image and set the positive detection threshold for a true source 1$\sigma$ above the absolute value of the brightest negative pixel.
    \item Test 3: Concatenated the calibrated $uv$ data of the target in multiple epochs using the AIPS task \textsc{DBCON} and imaged this new $uv$ database to find a significant detection (>6$\sigma$). Targets that are not moving but are faint can be detected using this technique.
\end{itemize}

\section{Results}\label{Section 4}
\begin{table}
  \centering
  \caption{Positions of the targets detected after the imaging process. Targets marked with a 'D' superscript indicate the positions are reported after concatenating data from two epochs. }
    \begin{tabular}{llll}
     Target & MJD & \multicolumn{2}{c}{Positions}  \\ \hline
            &     & \multicolumn{1}{c}{RA} & \multicolumn{1}{c}{Dec} \\ 
            &     & \multicolumn{1}{c}{h:m:s (error)} & \multicolumn{1}{c}{d:m:s (error)} \\    \hline
     A\romannumeral 1 & 57248 & 19:02:47.990403\phantom{1} (86)  & 05:29:21.37462 (20)\\
                      & 57382 & 19:02:47.990458\phantom{1} (13) & 05:29:21.37449 (34)\\
     A\romannumeral 3 & 57248 & 19:00:37.030793\phantom{1} (30) & 04:14:59.16581 (19)\\
     B\romannumeral 2 & 57173 & 18:53:21.381952\phantom{1} (12)   & -00:29:04.26976 (29) \\
                      & 57335 & 18:53:21.381867\phantom{1} (15)   & -00:29:04.26945 (33) \\
                      & 57418 & 18:53:21.381912\phantom{1} (10)   & -00:29:04.26961 (23) \\
     C\romannumeral 1 & 57173 & 18:52:24.294895\phantom{1} (10)    & -00:30:29.56350 (19) \\
                      & 57389 & 18:52:24.295672\phantom{1} (60)  & -00:30:29.56893 (30) \\
     D\romannumeral 1 & 57249 & 18:48:48.093142\phantom{1} (26) & -01:39:54.68671 (54) \\
                      & 57386 & 18:48:48.093298\phantom{1} (27) & -01:39:54.68980 (74) \\
     E\romannumeral 1 & 57174 & 18:44:51.4623927 (97)    & -02:20:05.80993 (17) \\
                      & 57362 & 18:44:51.4623623 (114)   & -02:20:05.80947 (17) \\
                      & 57422 & 18:44:51.4623300 (91)    & -02:20:05.80995 (18) \\
     E\romannumeral 2 & 57174 & 18:46:42.0492619 (80)    & -03:00:24.33092 (15) \\
                      & 57362 & 18:46:42.0492235 (120)    & -03:00:24.33027 (19) \\
                      & 57422 & 18:46:42.0492214 (97)    & -03:00:24.33044 (16) \\
     F\romannumeral 2 & 57398 & 18:46:14.072095\phantom{1} (24) & -03:11:40.53911 (40) \\
     G\romannumeral 2$^{\mathrm{D}}$ & 57398 & 18:45:04.298963\phantom{1} (45) & -03:22:49.68168 (95) \\
     H\romannumeral 1 & 57319 & 18:45:00.3479320 (95)    & -03:28:25.62092 (23) \\
                      & 57402 & 18:45:00.3477892 (94)    & -03:28:25.62246 (21) \\
     H\romannumeral 2 & 57319 & 18:44:47.2874434 (75)    & -03:59:36.46651 (18) \\
                      & 57402 & 18:44:47.2873573 (86)    & -03:59:36.46626 (20) \\
     I\romannumeral 1$^{\mathrm{D}}$ & 57215 & 18:41:33.288199\phantom{1} (14) & -04:24:33.09082 (25) \\
     I\romannumeral 2$^{\mathrm{D}}$ & 57391 & 18:38:26.363727\phantom{1} (18) & -05:48:35.13265 (27)\\
     M\romannumeral 1$^{\mathrm{D}}$ & 57244 & 18:34:33.047924\phantom{1} (20)& -08:15:26.96691 (32) \\
     \hline
    \end{tabular}%
\label{Postions_VLBA_survey}
\end{table}%

Results of the data reduction are summarised in Table \ref{Detections_VLBA_survey}, Table \ref{Postions_VLBA_survey} and Table \ref{tb_spec}. For the sources that were not detected and only went through the first order data reduction, we have reported their rms noise in Table \ref{Detections_VLBA_survey}. We detected (at 5$\sigma$ or greater significance) 14 out of the 33 targets in one or more epochs. We have discussed the possible reasons for non-detections of the 19 targets in Section \ref{non-detection-discussion}. Below we discuss the two targets that were detected and whose positions could be measured for a single epoch (A\romannumeral 3, F\romannumeral 2), followed by the eight targets that had detections and position measurements for multiple epochs (A\romannumeral 1, B\romannumeral 2, C\romannumeral 1, D\romannumeral 1, E\romannumeral 1, E\romannumeral 2, H\romannumeral 1\, and H\romannumeral 2). We also discuss the four targets that show reliable detections after concatenating data from two epochs (G\romannumeral 2, I\romannumeral 1, I\romannumeral 2\, and M\romannumeral 1). 

\subsection{Single epoch detections}\label{single_ep_results}
We could measure the positions of two targets (A\romannumeral 3\, and F\romannumeral 2) for only a single epoch (see Table \ref{Detections_VLBA_survey}, \ref{Postions_VLBA_survey} and \ref{tb_spec} for details). The contour maps of the targets are shown in Figure \ref{A3_F2_contour}. A\romannumeral 3\, was only successfully observed for one epoch. This target was part of the observations in which we required the multi-phase centre correlation capability of DiFX. Due to an error, the data were correlated only at the pointing centre and not at the target positions in the second epoch of the observations for these targets. Therefore, the time and bandwidth smearing at the positions of the targets were too large to yield a useful data set. F\romannumeral 2\phantom{1}was observed for three epochs but was only detected in the first epoch. The source appears marginally resolved and we used {\tt{uvtaper}} to down-weight the baselines longer than 20\,M$\lambda$. 

\begin{table*}
\caption{Source sizes ($\theta_{\mathrm{maj}}$ and $\theta_{\mathrm{min}}$), brightness temperatures (T$_{\mathrm{b}}$), and the THOR and GLOSTAR spectral indices of the sources detected in our VLBA sample. We have reported the upper limits of the major and minor axes for compact sources, and thus could estimate the lower limit of the brightness temperature of such sources.  }
\begin{tabular}{lccccccccccc}
\hline
Name & \multicolumn{2}{c}{Size of source} & \multicolumn{1}{c}{}       & \multicolumn{2}{c}{Size of source} & \multicolumn{1}{c}{}   & \multicolumn{2}{c}{Size of source} & \multicolumn{1}{c}{}   & \multicolumn{2}{c}{Spectral index}  \\
\hline
& \multicolumn{3}{c}{Epoch 1}  & \multicolumn{3}{c}{Epoch 2}  & \multicolumn{3}{c}{Epoch 3} & \multicolumn{1}{c}{THOR} & \multicolumn{1}{c}{GLOSTAR} \\
 & $\theta_{\mathrm{maj}}$ & $\theta_{\mathrm{min}}$ & T$_{\mathrm{b}}$ & $\theta_{\mathrm{maj}}$ & $\theta_{\mathrm{min}}$ & T$_{\mathrm{b}}$ & $\theta_{\mathrm{maj}}$ & $\theta_{\mathrm{min}}$ & T$_{\mathrm{b}}$ &                          &                             \\
 & (mas) & (mas) & ($\times$10$^{6}$\,K) & (mas) & (mas) & ($\times$10$^{6}$\,K) & (mas) & (mas) & ($\times$10$^{6}$\,K) &    & \\
     \hline
G39.1105-0.0160 (A\romannumeral 1)  & \textless{}2 & \textless{}2 & \textgreater{}3 & \textless{}5  & \textless{}1  & \textgreater{}3 & - & - & - & $-$0.15$\pm$0.20 & - \\
G37.7596-0.1001 (A\romannumeral 3) & 36$\pm$0.1 & 8.3$\pm$0.5 & 3 & - & - & - & - & - & - & - & - \\
G32.7193-0.6477 (B\romannumeral 2) & \textless{}4 & \textless{}2 & \textgreater{}1.2 & \textless{}5 & \textless{}3 & \textgreater{}0.7  & \textless{}4  & \textless{}3 & \textgreater{}1  & $-$1.05$\pm$0.2 & $-$0.65$\pm$0.11 \\
G32.5898-0.4468 (C\romannumeral 1) & \textless{}3 & \textless{}2 & \textgreater{}3.3 & \textless{}4  & \textless{}2 & \textgreater{}2.4  & -  & - & - & $-$0.67$\pm$0.3 & $-$0.12$\pm$0.13 \\
G31.1494-0.1727 (D\romannumeral 1) & \textless{}5 & \textless{}5 & \textgreater{}0.5 & \textless{}4 & \textless{}5 & \textgreater{}0.7 & - & - & - & - & \phantom{1}1.11$\pm$0.18  \\
G30.1038+0.3984 (E\romannumeral 1)  & 4$\pm$1 & 3$\pm$1 & 1.6 & 5$\pm$1 & 3$\pm$1 & 1.4 & 5$\pm$1 & 4$\pm$1 & 1.1 & $-$0.69$\pm$0.4 & $-$0.69$\pm$0.12 \\
G29.7161-0.3178 (E\romannumeral 2) & 9$\pm$1 & 7$\pm$1 & 0.9 & 9$\pm$1 & 7$\pm$1 & 1.1 & 8$\pm$1 & 8$\pm$1 & 1 &         - & - \\
G29.4959-0.3000 (F\romannumeral 2) & 6$\pm$2 & 5$\pm$2 & 0.7 & - & - &- & - & - & - & \phantom{12}1.13$\pm$0.67 & -1.05$\pm$0.9 \\
G29.1978-0.1268 (G\romannumeral 2) & 13$\pm$2 & 3$\pm$1 & 0.2 & - & - & - & - & - & - & \phantom{1}-1.4$\pm$1.2 & \phantom{11}0.18$\pm$0.18\\
G29.1075-0.1546 (H\romannumeral 1) & \textless{}3 & \textless{}2 & \textgreater{}2.7 & \textless{}3 & \textless{}2 & \textgreater{}2.5 & -& - &- &\phantom{11}0.26$\pm$0.23 & \phantom{1}-0.54$\pm$0.06\\
G28.6204-0.3436 (H\romannumeral 2)  & \textless{}3 & \textless{}2 & \textgreater{}3.2 & \textless{}3 & \textless{}2 & \textgreater{}2.4 & - & - & - & \phantom{12}-1.0$\pm$0.14 & \phantom{12}-1.4$\pm$0.11\\
G27.8821+0.1834 (I\romannumeral 1) & 8$\pm$1 & 7$\pm$1 & 0.7 &  - &  - & - & - & - & - & \phantom{1}-0.11$\pm$0.28 & - \\
G26.2818+0.2312 (I\romannumeral 2) & 12$\pm$1 & 9$\pm$1 & 0.7 & - & - & - & - & - & -  & \phantom{12}0.00$\pm$0.13 & -\\
G23.6644-0.0372 (M\romannumeral 1) & 14$\pm$1  & 13$\pm$1 & 1.6 & - & - & - & -  &  -  &  -  & \phantom{12}0.63$\pm$0.46 &  -     \\        
\hline
\end{tabular}
\label{tb_spec}
\end{table*}

\subsection{Multiple epoch detections}
\begin{figure*}
\begin{center}
\begin{tabular}{cc}
\phantom{123456} A\romannumeral 1 & \phantom{123456} B\romannumeral 2 \\
\includegraphics[width=8cm,height=7cm,angle=0]{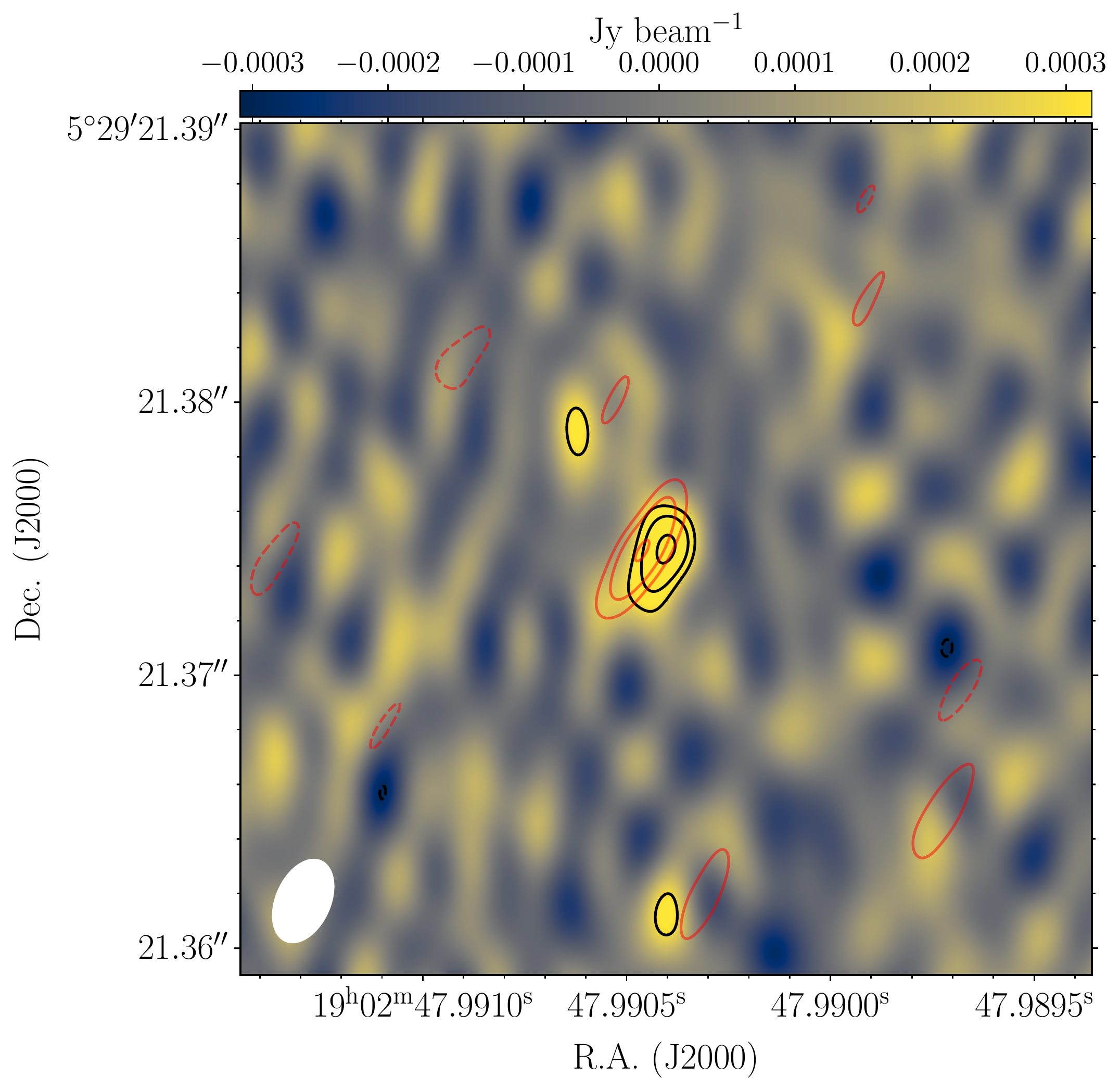} &
\includegraphics[width=8cm,height=7cm,angle=0]{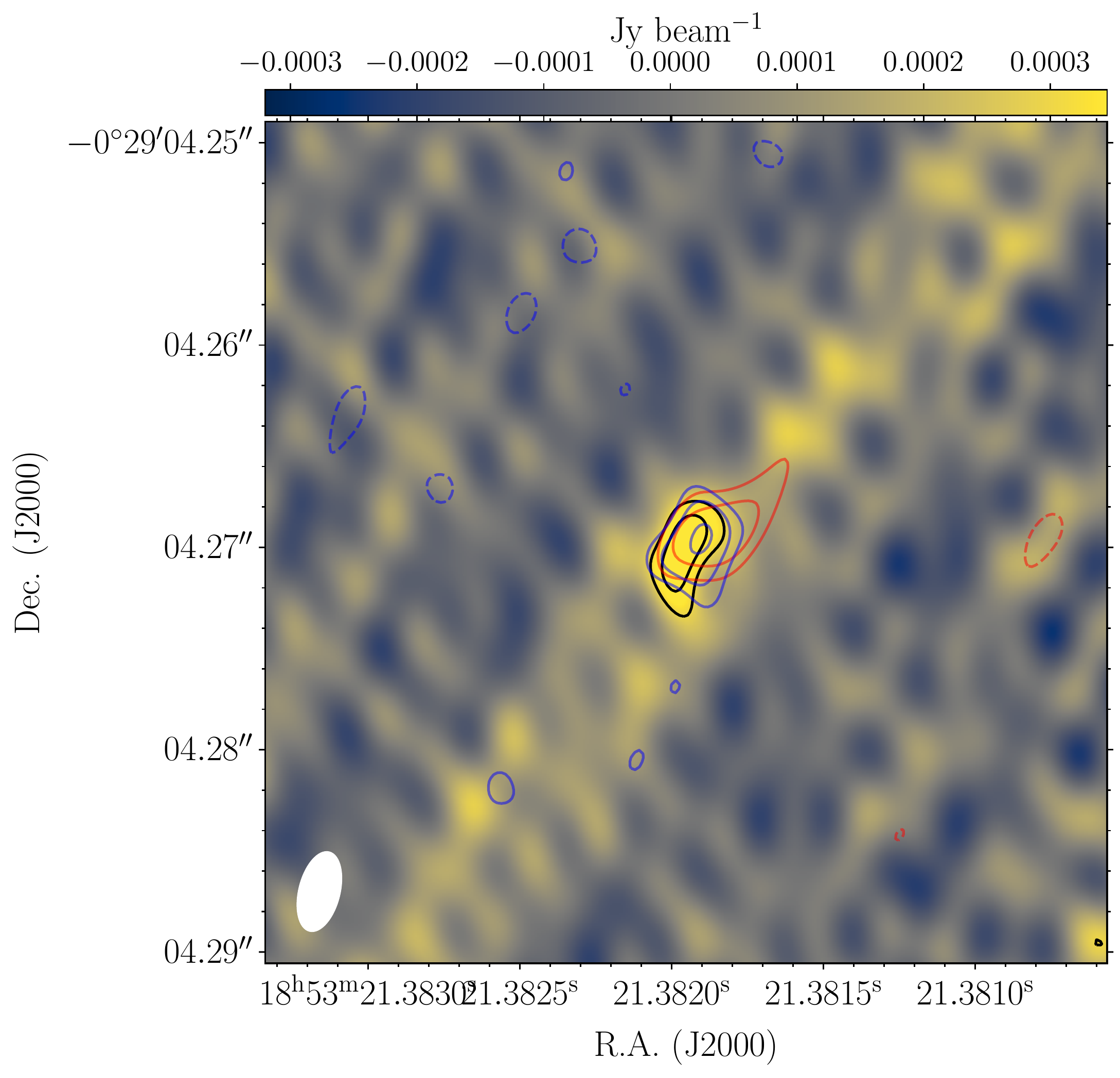}  \\
\phantom{123456765} C\romannumeral 1 & \phantom{123456765} D\romannumeral 1 \\
\includegraphics[width=8cm,height=7cm,angle=0]{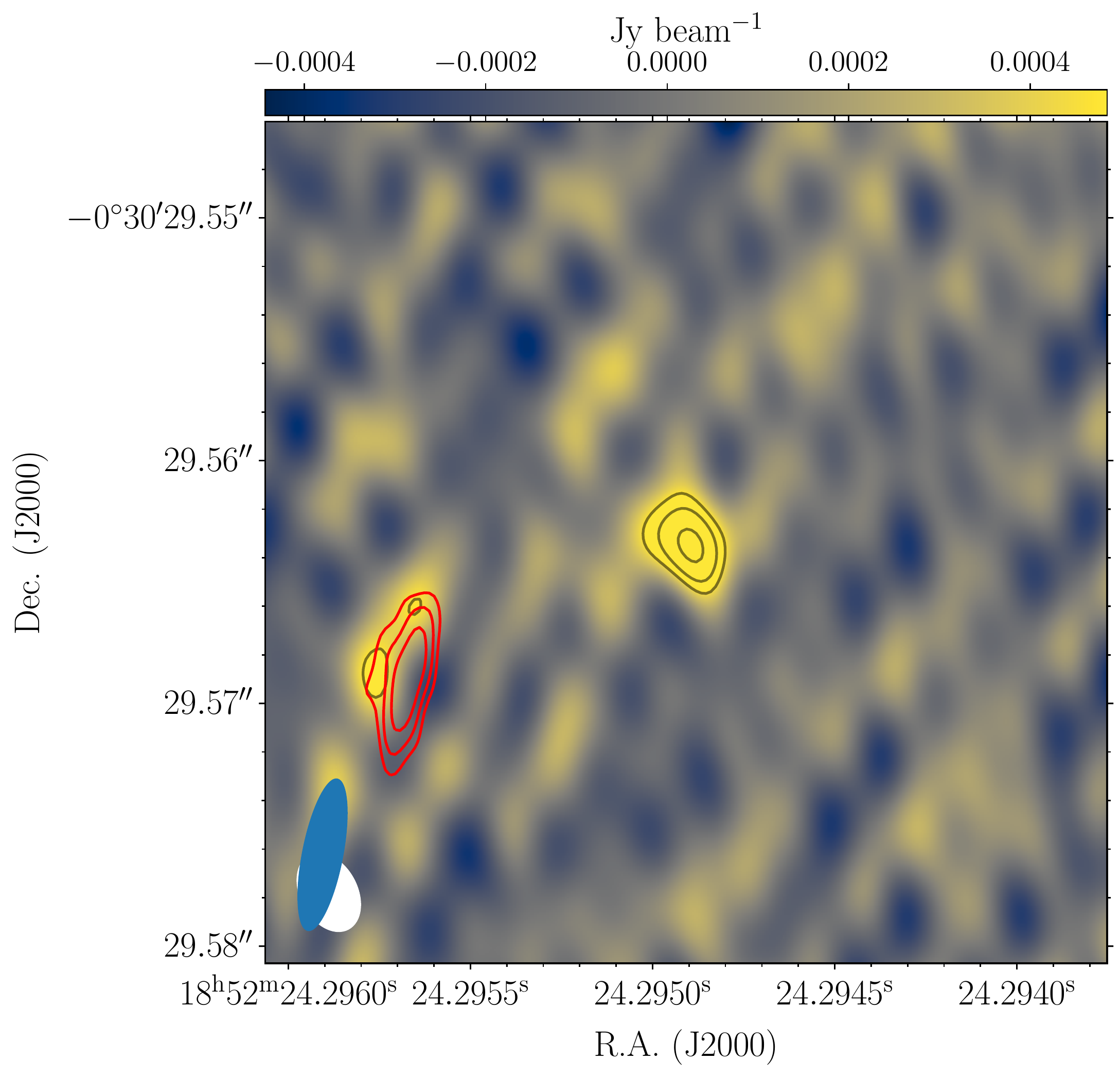} & 
\includegraphics[width=8cm,height=7cm,angle=0]{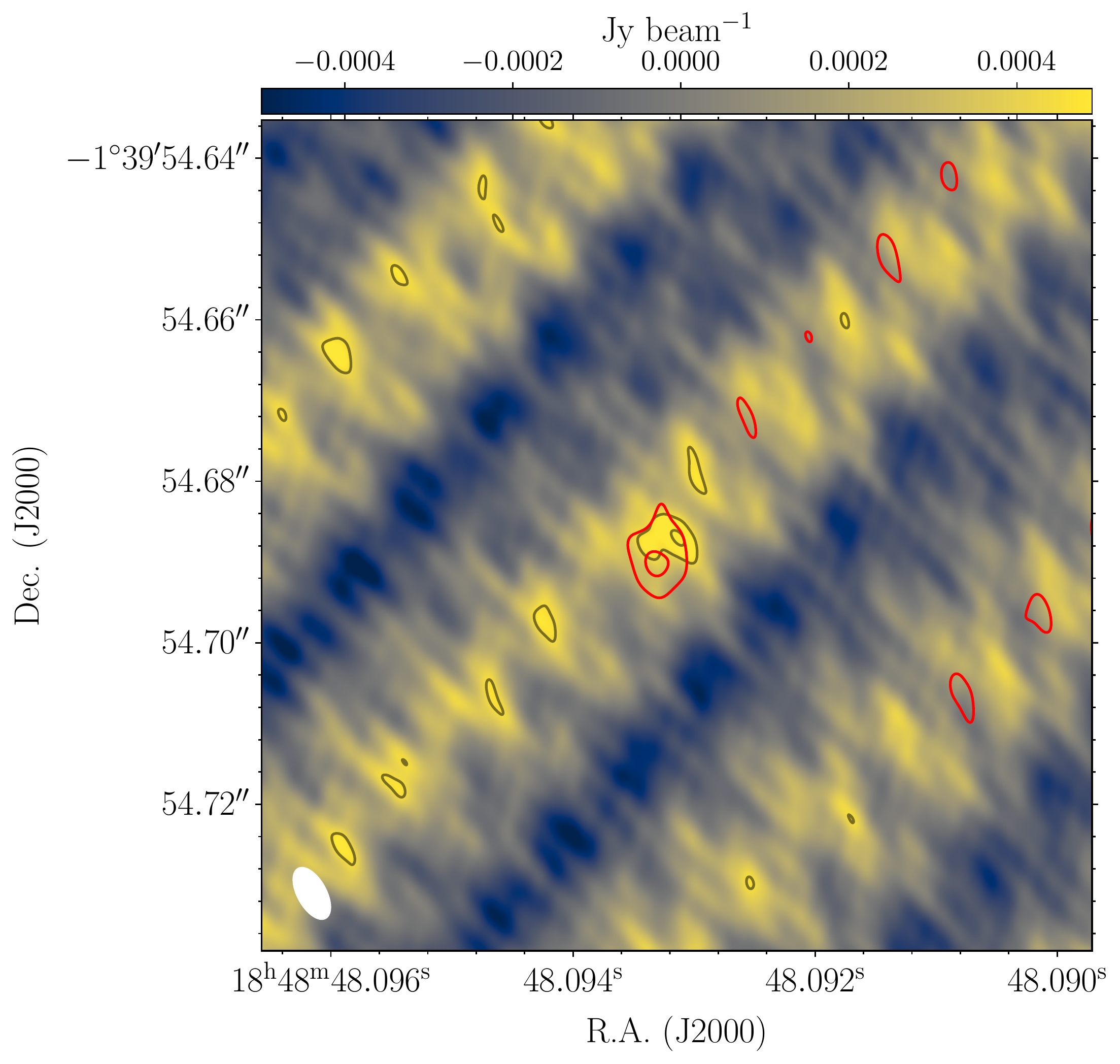} \\
\phantom{123456765} H\romannumeral 1 & \phantom{123456765} H\romannumeral 2 \\
\includegraphics[width=8cm,height=7cm,angle=0]{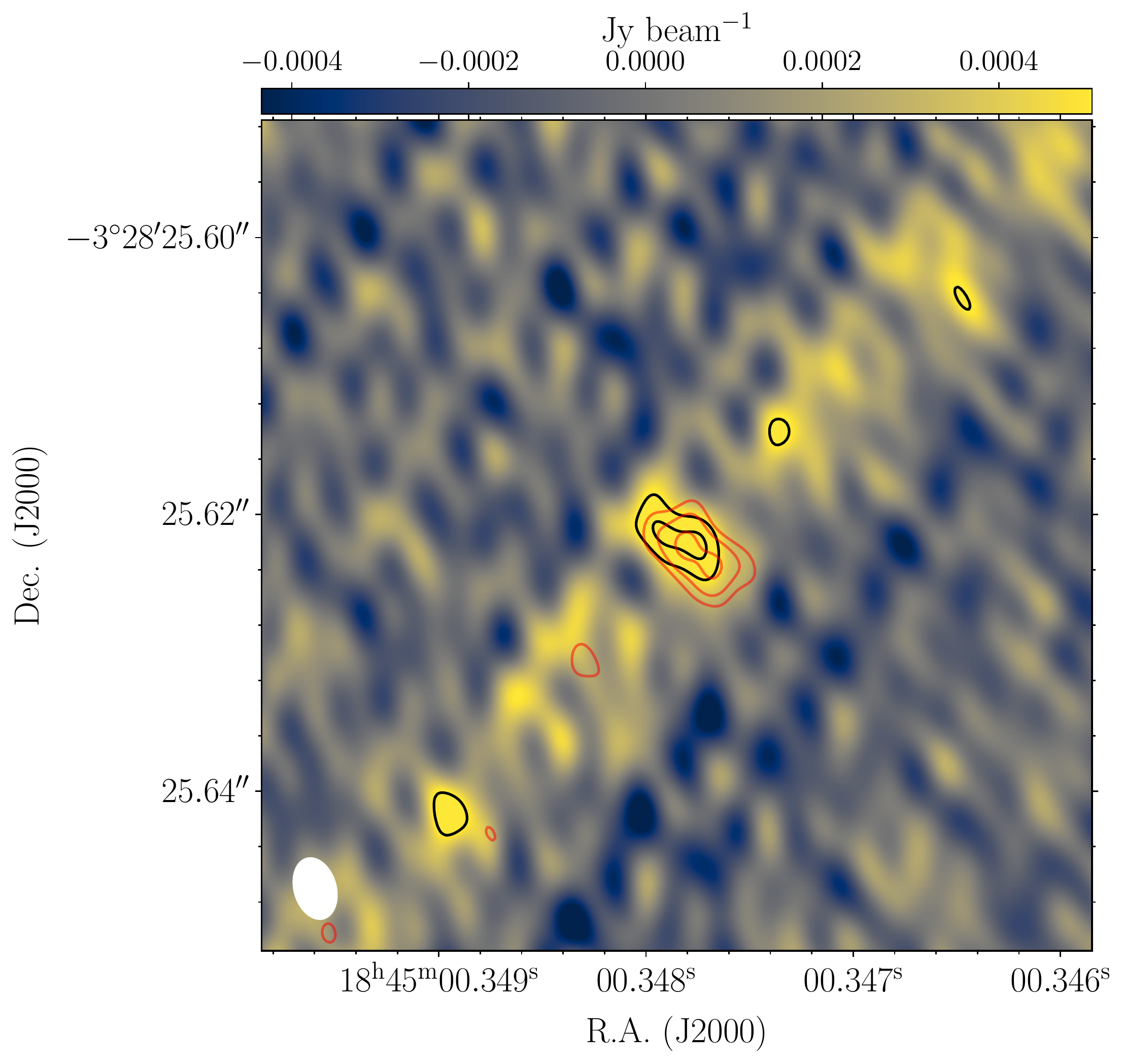} & 
\includegraphics[width=8cm,height=7cm,angle=0]{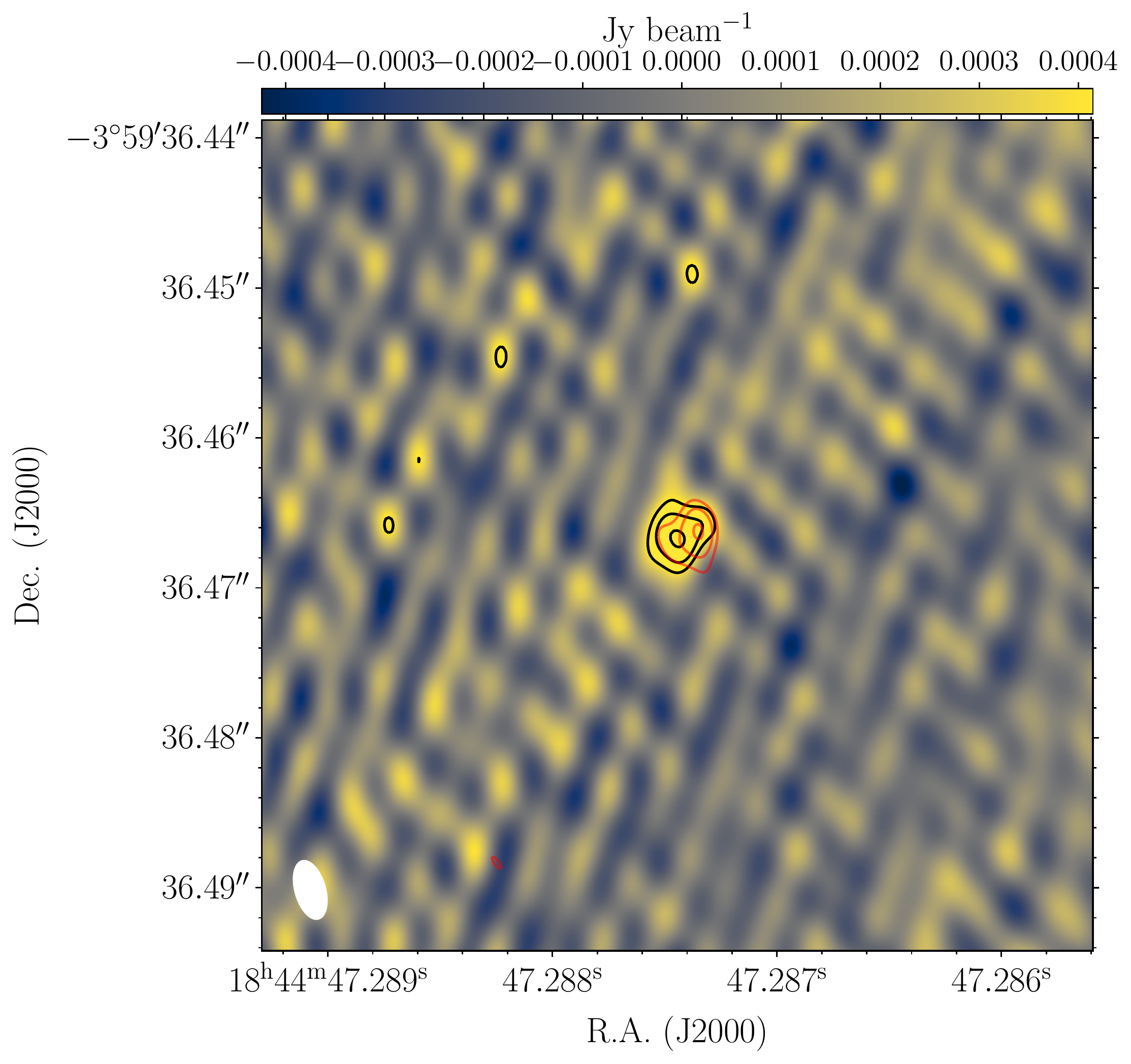} \\
\end{tabular}
\end{center}
\caption{ \label{multiple_epoch_contours} Image of the Gaussian elliptical model fit to the six multiple epoch detections of compact targets. The targets are G39.1105-0.0160 (A\romannumeral 1; \textit{top-left}), G32.7193-0.6477 (B\romannumeral 2; \textit{top-right}), G32.5898-0.4468 (C\romannumeral 1; \textit{middle left}), G31.1494-0.1727 (D\romannumeral 1; \textit{middle-right}), G29.1075-0.1546 (H\romannumeral 1; \textit{bottom-left}) and G28.6204-0.3436 (H\romannumeral 2; \textit{bottom-right}). The white ellipses in the left-bottom corner of each image represents the beam size and shape, and the blue ellipse in C\romannumeral 1\, is the beam of the second epoch as it was markedly different from the first epoch. The contours levels are in steps of rms of each image with increments of sqrt(2), with the lowest contour at 3$\times$rms. Black contours are first epoch detections, red contours are second epoch detections and blue contours are third epoch detections. }
\end{figure*}

There were seven targets that were detected with a significance of $\geq 6\sigma$ in multiple epochs, and one target (D\romannumeral 1\,) with $\sim$5$\sigma$ detections in both the epochs it was observed in. We imaged all the targets with all the antennas, as well as with down-weighting the data from the longer baselines to identify any targets that are resolved. We find that G30.1038+0.3984 (E\romannumeral 1) and G29.7161-0.3178 (E\romannumeral 2) show extended structure, whereas all the other targets (A\romannumeral 1, B\romannumeral 2, C\romannumeral 1, D\romannumeral 1, H\romannumeral 1\, and H\romannumeral 2) appear compact. The contour maps of all the targets with detections in multiple epochs are shown in Figure \ref{multiple_epoch_contours}. The details of these detections are reported in Table \ref{Detections_VLBA_survey}, \ref{Postions_VLBA_survey} and \ref{tb_spec}. The data reduction details of some of the targets are given below. 
\subsubsection{G32.7193-0.6477 (B\romannumeral 2)}
B\romannumeral 2\, was detected with a significance of 5.7$\sigma$ in the first epoch, and with a significance of $>$6$\sigma$ in the second and the third epoch. The target was imaged without Mauna Kea in all three epochs as timing errors were reported in the first and second epoch observations. Although the detection in the first epoch is below our detection threshold of 6$\sigma$, the presence of a $>$6$\sigma$ source in the second and third epoch within a few pixels of the first epoch detection and with a similar flux density suggests that the first epoch is a reliable detection. We therefore include all three epochs' positions for analysis.
\subsubsection{G31.1494-0.1727 (D\romannumeral 1)}
D\romannumeral 1\, was detected in both the epochs it was observed in with 5--6$\sigma$ significance when imaged with a robustness parameter of 5. The flux density of the false positives in both these observations is within 1$\sigma$ of the detection and hence these detections fail Test 1. On the other hand, these detections clear our Test 2 and could be astrophysical sources. We also note that the D\romannumeral 1\, detections are separated by four months and are within a few pixels of each other with similar flux density. It is much less likely that two independent images will have $\geq 5\sigma$ noise peaks within a few pixels of each other. Hence, we treat the detections of D\romannumeral 1\, as true detections. We flagged Mauna Kea and St. Croix for both datasets as the phase solutions at both these antennas were very sparse. 

\subsubsection{G30.1038+0.3984 (E\romannumeral 1)}
E\romannumeral 1\, was detected in all the three epochs with a $>$15$\sigma$ significance when imaged with a robustness parameter of 0 and all the antennas. The observations have no data from St. Croix and Mauna Kea. The target is resolved; however, their integral intensity and angular scales do not change when using the Gaussian $uv$ tapering of 20\,M$\lambda$ and when using all the antennas, so we report the positions and flux densities of the target using all the antennas. 

\subsubsection{G29.7161-0.3178 (E\romannumeral 2)}
E\romannumeral 2\, was detected in all three epochs when imaged with a robustness parameter of 0 and all the antennas. The observations have no data from St. Croix and Mauna Kea. The positions, angular scales and integral intensity of the target does not change when imaged with a Gaussian $uv$ taper of 20\,M$\lambda$, hence we report the position and angular scale of the target when imaged with all the antennas.

\begin{figure*}
\begin{center}
\begin{tabular}{cc}
\phantom{123456} E\romannumeral 1 & \phantom{123456} E\romannumeral 2 \\
\includegraphics[width=8cm,height=7cm,angle=0]{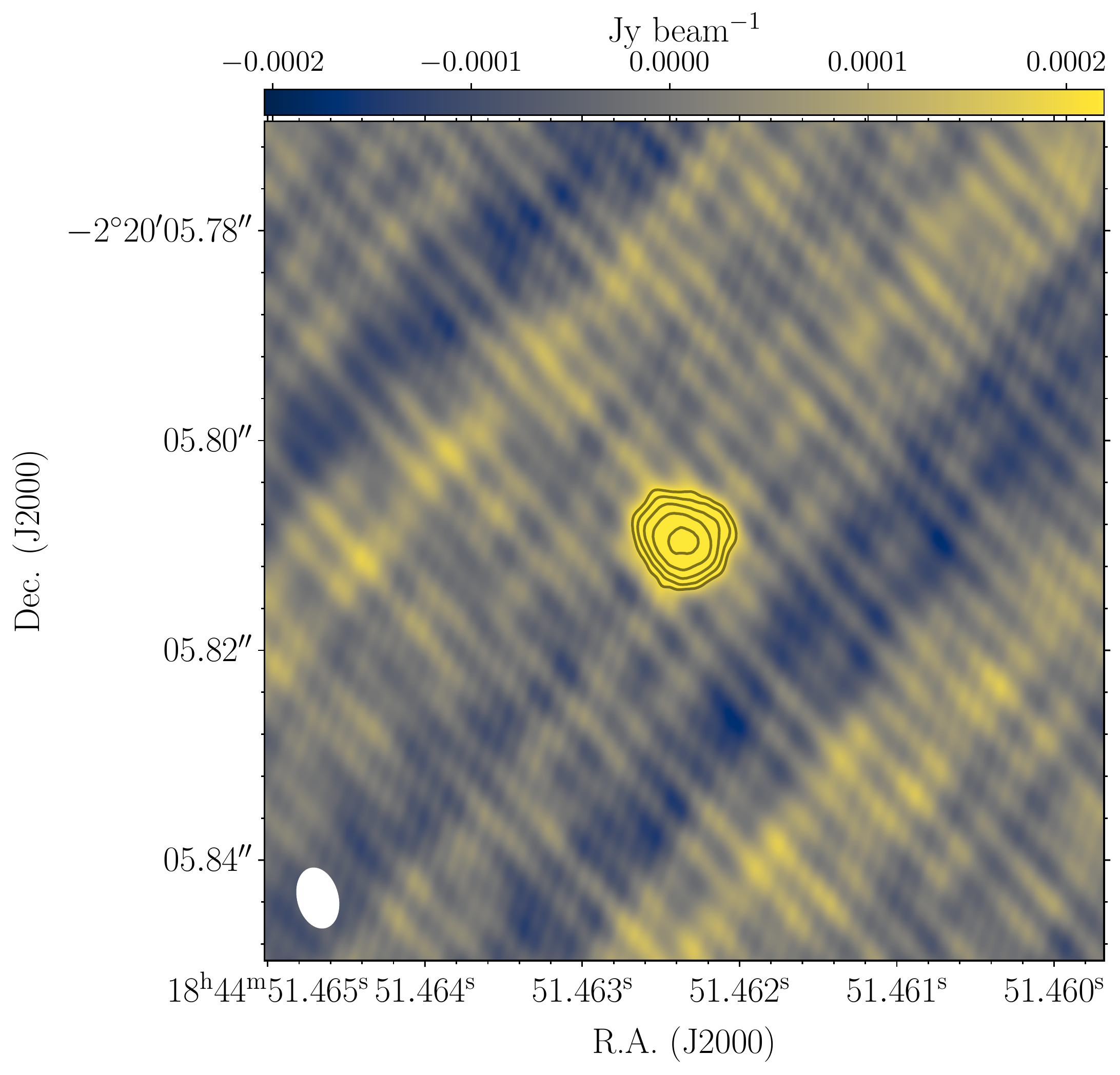} &
\includegraphics[width=8cm,height=7cm,angle=0]{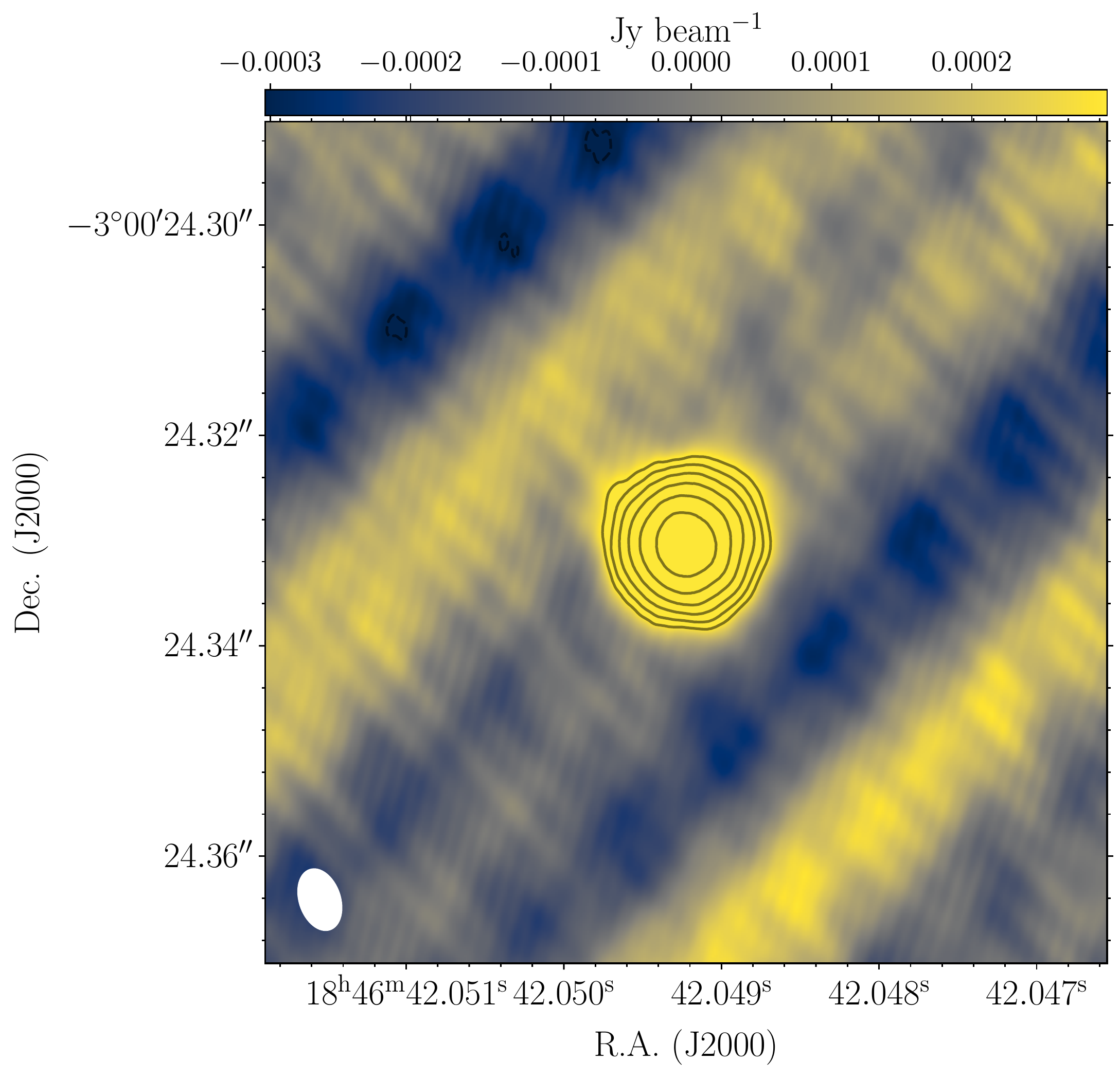}  \\
\end{tabular}
\end{center}
\caption{ \label{concat_contour} Image and contours of the Gaussian elliptical model fit to the G30.1038+0.3984 (E\romannumeral 1; \textit{left}) and G29.7161-0.3178 (E\romannumeral 2; \textit{right}) as detected in the second epoch. The targets showed consistent size and flux density between the epochs, thus we only show the contour map of both these targets for a single epoch (the second epoch). The white ellipses in the left-bottom corner of each image represents the beam size and shape. The contours levels are in steps of rms of each image with increments of sqrt(2), with the lowest contour at 3$\times$rms. }
\end{figure*}

\subsection{Detections in concatenated data}\label{dbcon_results}
\begin{figure*}
\begin{center}
\begin{tabular}{cc}
\phantom{123456} G\romannumeral 2 & \phantom{123456} I\romannumeral 1 \\
\includegraphics[width=8cm,height=7cm,angle=0]{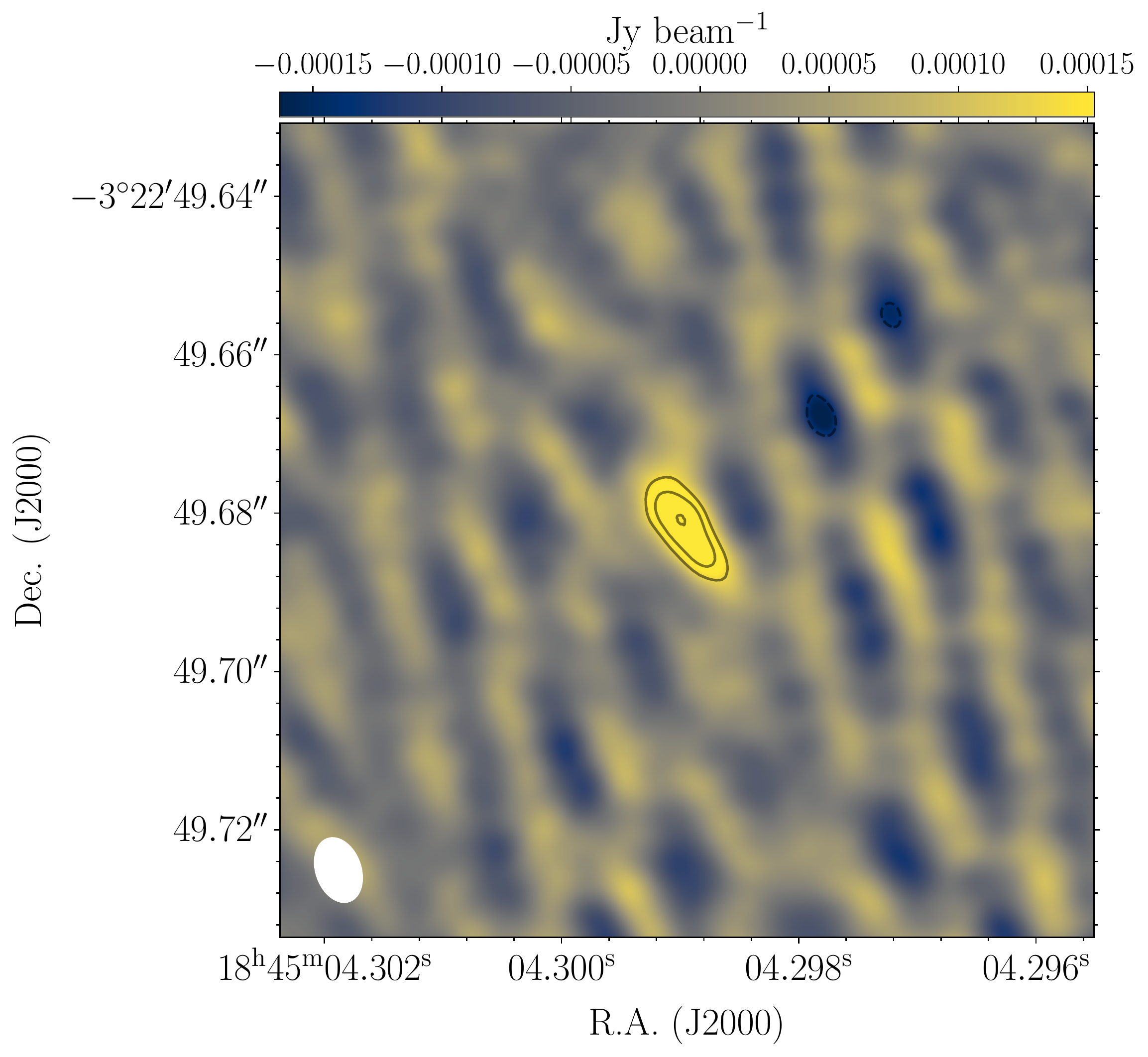} &
\includegraphics[width=8cm,height=7cm,angle=0]{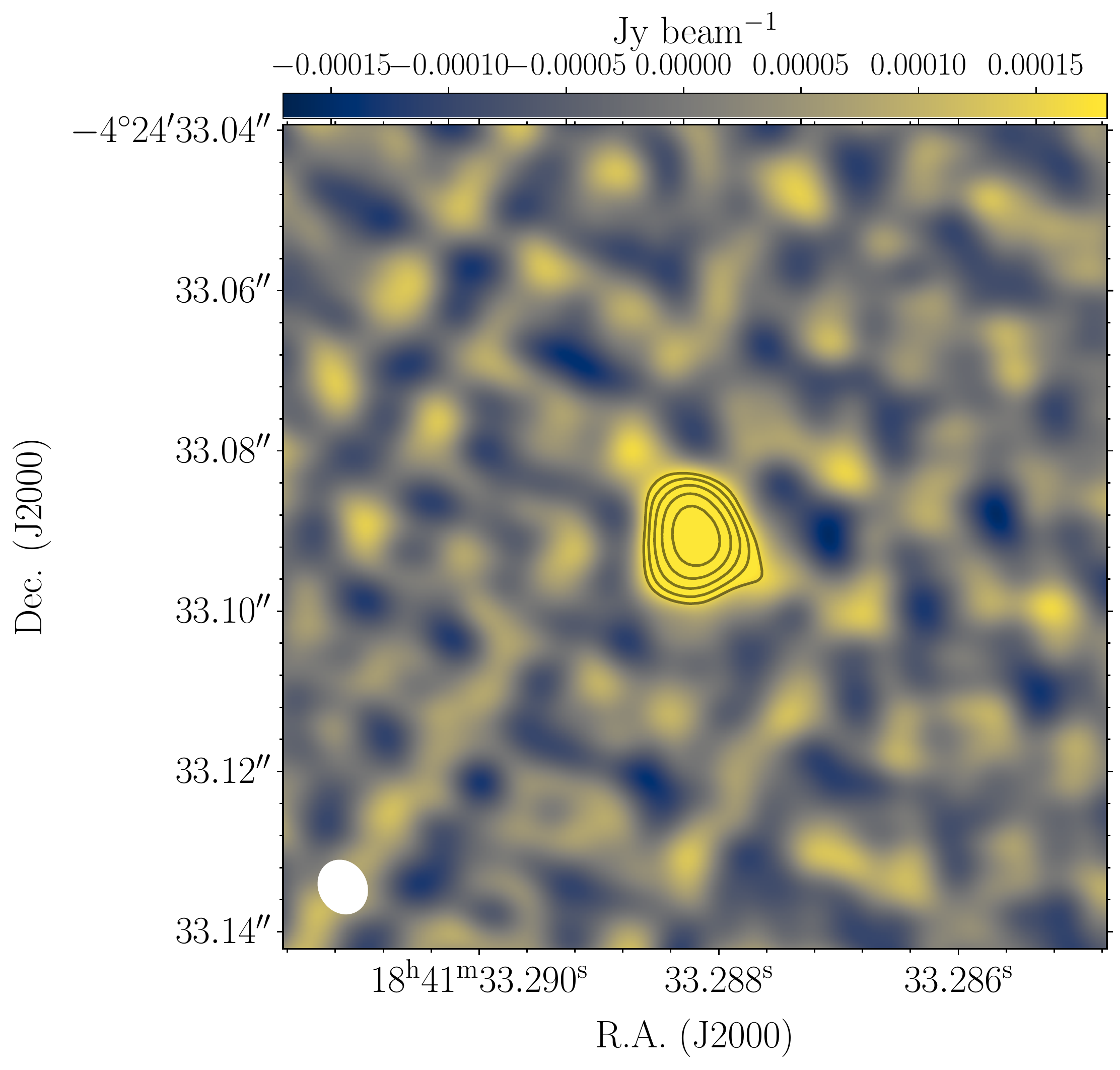}  \\
\phantom{123456} I\romannumeral 2 & \phantom{123456} M\romannumeral 1 \\
\includegraphics[width=8cm,height=7cm,angle=0]{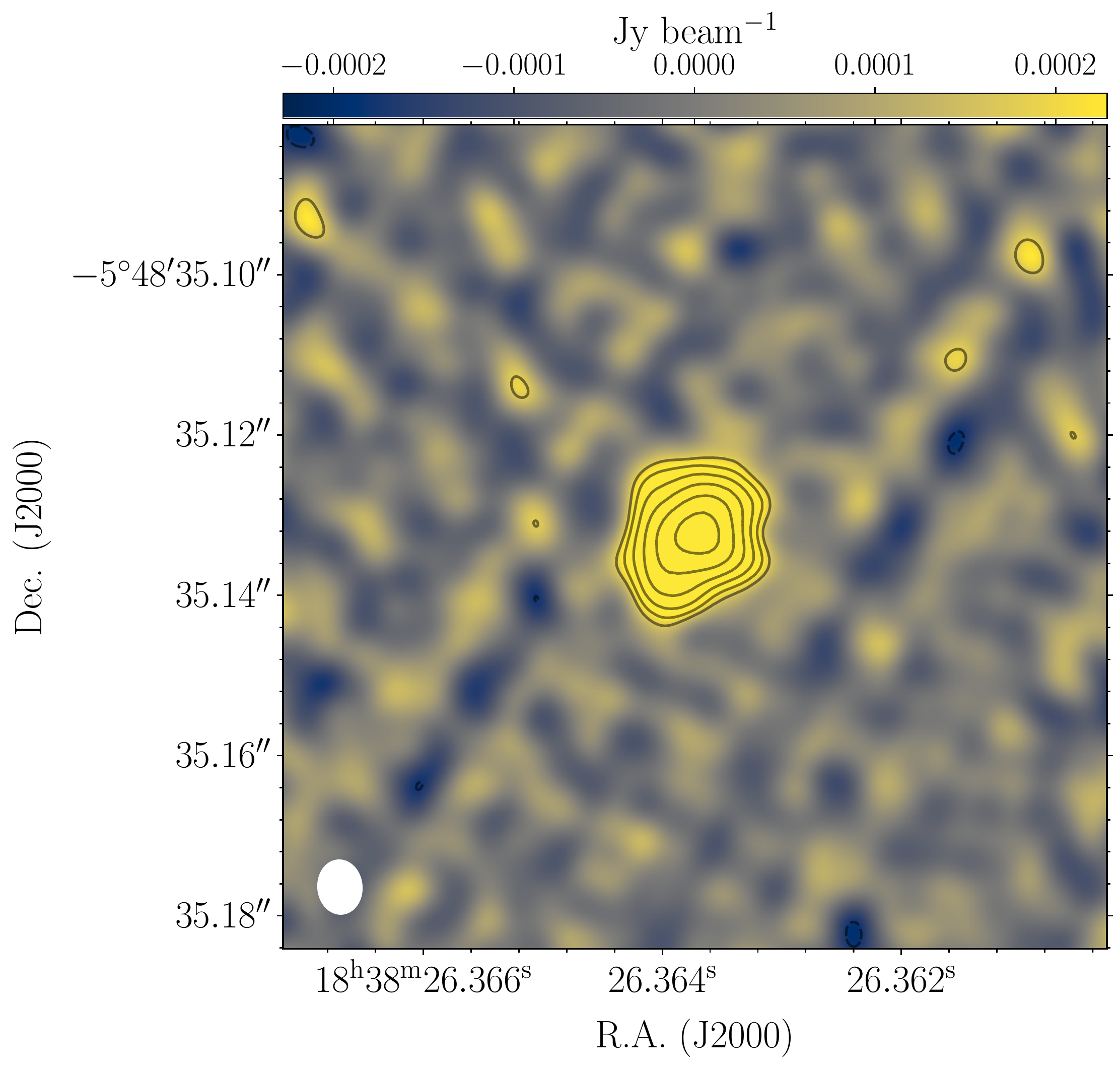} &
\includegraphics[width=8cm,height=7cm,angle=0]{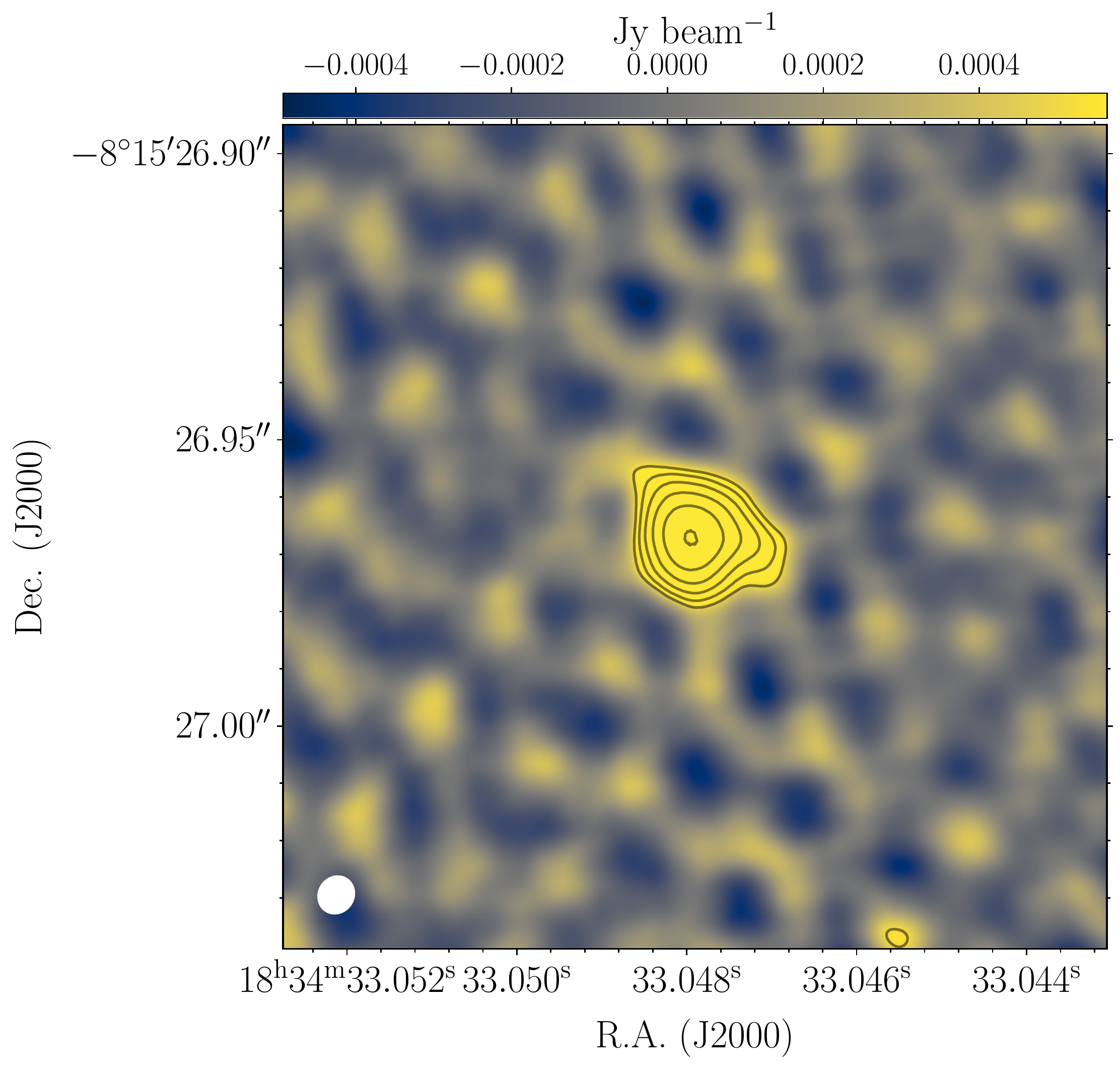}  \\
\end{tabular}
\end{center}
\caption{ \label{concat_contour_2} Image and contours of the Gaussian elliptical model fit to the sources with detections reported after concatenating data. The sources are G29.6051-0.8590 (G\romannumeral 2; \textit{top-left}), G27.8821+0.1834 (I\romannumeral 1; \textit{top-right}), G26.2818+0.2312 (I\romannumeral 2; \textit{bottom-left}) and G23.6644-0.0372 (M\romannumeral 1; \textit{bottom-right}). The white ellipses in the left-bottom corner of each image represents the beam size and shape. The contours levels represent rms of each image with increments of sqrt(2), with the lowest contour at 3$\times$rms.  }
\end{figure*}

We report detections of four targets after concatenating data from multiple epochs due to the reasons stated with the targets below. The detections are reported in Table \ref{Detections_VLBA_survey}, \ref{Postions_VLBA_survey} and \ref{tb_spec}. The contour maps are shown in Figure \ref{concat_contour} and Figure \ref{concat_contour_2}.
\subsubsection{G29.1978-0.1268 (G\romannumeral 2)}
We imaged G\romannumeral 2\, by applying a Gaussian uvtaper with a $uv$-width of 20\,M$\lambda$. We could see fringe patterns in both the epochs but could not conclusively find the location of the target. We concatenated the data of both the epochs to increase the $uv$-coverage, and then imaged again by setting a robustness parameter of 5 (natural weighting) and using the same Gaussian taper as before. We detected the target with a peak intensity significance of 6$\sigma$ after concatenating the data and the target appeared resolved. 

\subsubsection{G27.8821+0.1834 (I\romannumeral 1)}
I\romannumeral 1\, was observed for two epochs. The Pie Town dish did not observe in the second epoch. There is a bright fringe pattern detected in both epochs, although the location of the brightest pixel shifts depending on the antennas being used for imaging. Other than being the result of poor $uv$-coverage, this could also suggest that the target is resolved and so we imaged both the epochs by $uv$-tapering the longer baselines (Gaussian taper with a $uv$-width of 20\,M$\lambda$). In the first epoch, we detect a target with a peak intensity of 16$\sigma$ peak intensity (848$\pm$50\,$\upmu$Jy\,beam$^{-1}$) and an integral intensity of 2.2$\pm$0.2\,mJy, and the sizes of the major and minor axes of the source are 7$\pm$1\,mas and 6$\pm$1\,mas. The map of the second epoch of the target shows a 7$\sigma$ peak intensity (640$\pm$90\,$\upmu$Jy\,beam$^{-1}$) source with an integral intensity of  800$\pm$170\,$\upmu$Jy, and a major and minor axis size of 3$\pm$3\,mas and 0.2$^{+5.0}_{-0.2}$\,mas, respectively. There is an offset of 21\,mas in Declination in the centroid of the targets detected in the two epochs that are separated by 6 months. The size of the source is also inconsistent between the two epochs. Such a large implied motion in the Declination direction, a $<$6$\sigma$ significance of the integral intensity in the second epoch that is also half the value of the first epoch, and lack of an important antenna that contributes to the small baseline coverage (Pie Town) suggests that the second epoch detection of the target is not reliable. We thus concatenated the data for both the epochs to improve the $uv$-coverage and detect the source with a 18$\sigma$ significance. An elliptical Gaussian fit to the target in the $uv$-plane shows that the target has a major and minor axis of 8.4$\pm$1.0\,mas and 6.5$\pm$1.0\,mas, respectively, which is in agreement with the size of the source detected in the first epoch.
\subsubsection{G26.2818+0.2312 (I\romannumeral 2)}
I\romannumeral 2\, was observed for two epochs, and there is a bright fringe pattern seen in both epochs when imaged by down-weighting the longer baselines (uvtaper 20\,M$\lambda$) and with natural weighting (robustness parameter of 5). The exact position of the target is unclear in the second epoch with the brightest pixel moving in the image map depending on the baseline cutoff being used. We thus concatenate the data of the two epochs to increase the $uv$-coverage. The detection significance of the target increases from 17$\sigma$ (1.20$\pm$0.07\,mJy\,beam$^{-1}$) in the first epoch to 19$\sigma$ (1.20$\pm$0.05\,mJy\,beam$^{-1}$) in the concatenated data. The target is clearly resolved and an elliptical Gaussian model fit to the target gives a major axis of 11$\pm$1\,mas and minor axis of 9$\pm$1\,mas, which is also consistent with the size of the target as seen in the first epoch.

\subsubsection{G23.6644-0.0372 (M\romannumeral 1)}
M\romannumeral 1\, was observed for two epochs separated by 5 months. \citet{Yang2022}, Y22 hereon, had observed this source in 2010 using the European VLBI Network (EVN) and found that the source showed an extended structure at EVN resolution. We thus imaged the target only with the shortest baselines using a Gaussian taper and down-weighting baselines longer than 20\,M$\lambda$. The second epoch of observations had sparse $uv$-coverage as compared to the first epoch because the Los Alamos and Owens Valley dishes failed to take data for a large part of the experiment. We thus concatenate the two epochs of observations to improve $uv$-coverage and detect a resolved source.

\section{Analysis}\label{Section 5}
We use the positions measured in Section \ref{Section 4} to determine the proper motions of the five targets with detections in two epochs, and proper motion and parallax for one target that was detected in three epochs. We note that two targets (E\romannumeral 1\, and E\romannumeral 2) with detections in three epochs are seen as extended structures, and thus we are not certain that the centroid position being measured for these targets at every epoch is of the same component of the extended target. The proper motion in Right Ascension (RA; $\upmu_{\alpha}\mathrm{cos}\delta$), in Declination (Dec; $\upmu_{\delta}$), the parallax ($\pi$) and the reference position (RA$_{0}$, Dec$_{0}$) of the targets can be derived by fitting the measured position as a function of time \citep[e.g.][]{Loinard2007}. The time midway between the first and last observation of every target was used as the reference date. The input errors on the measured positions in RA and Dec were the statistical and systematic errors added in quadrature. The offset in the check source (the calibrator scans treated as a check source) position was smaller in magnitude than the more conservative estimate from the target-phase calibrator throw \citep{Pradel2006}. Since we are using calibrator scans as check source scans, and systematic error scales as calibrator-target throw, the check source offset are a lower limit to the systematic errors. So in all cases, the more conservative systematic error calculated by the relation from \citet{Pradel2006} was used. We used a Bayesian approach to determine $\upmu_{\alpha}\mathrm{cos}\delta$, $\upmu_{\delta}$ and $\pi$ using the equations that describe the motion of the target in the plane of the sky and using the measured positions of the targets as the data.\par
\begin{figure*}
\begin{center}
\begin{tabular}{cc}
\includegraphics[width=9cm,height=7cm,angle=0]{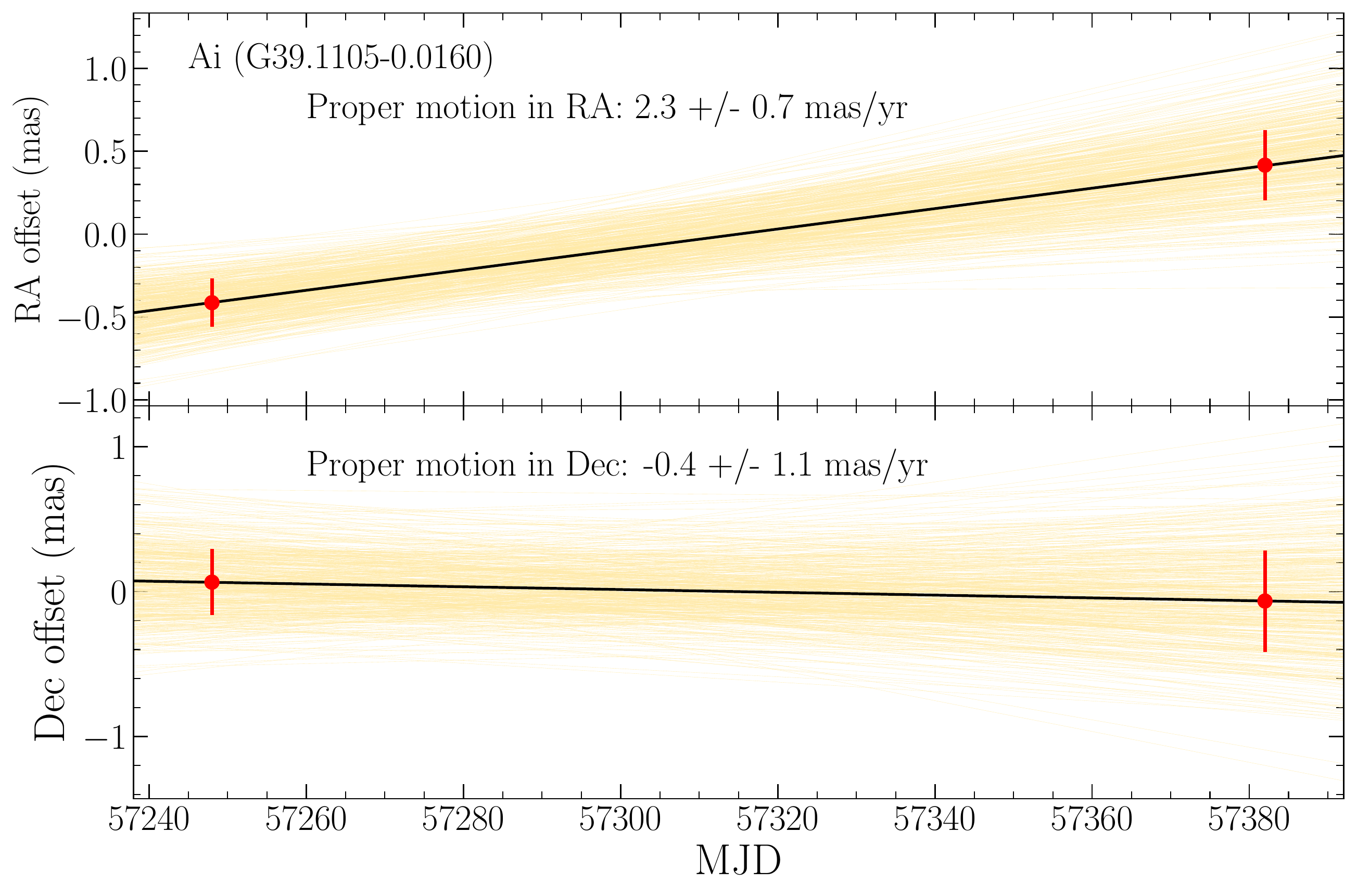} &
\includegraphics[width=9cm,height=7cm,angle=0]{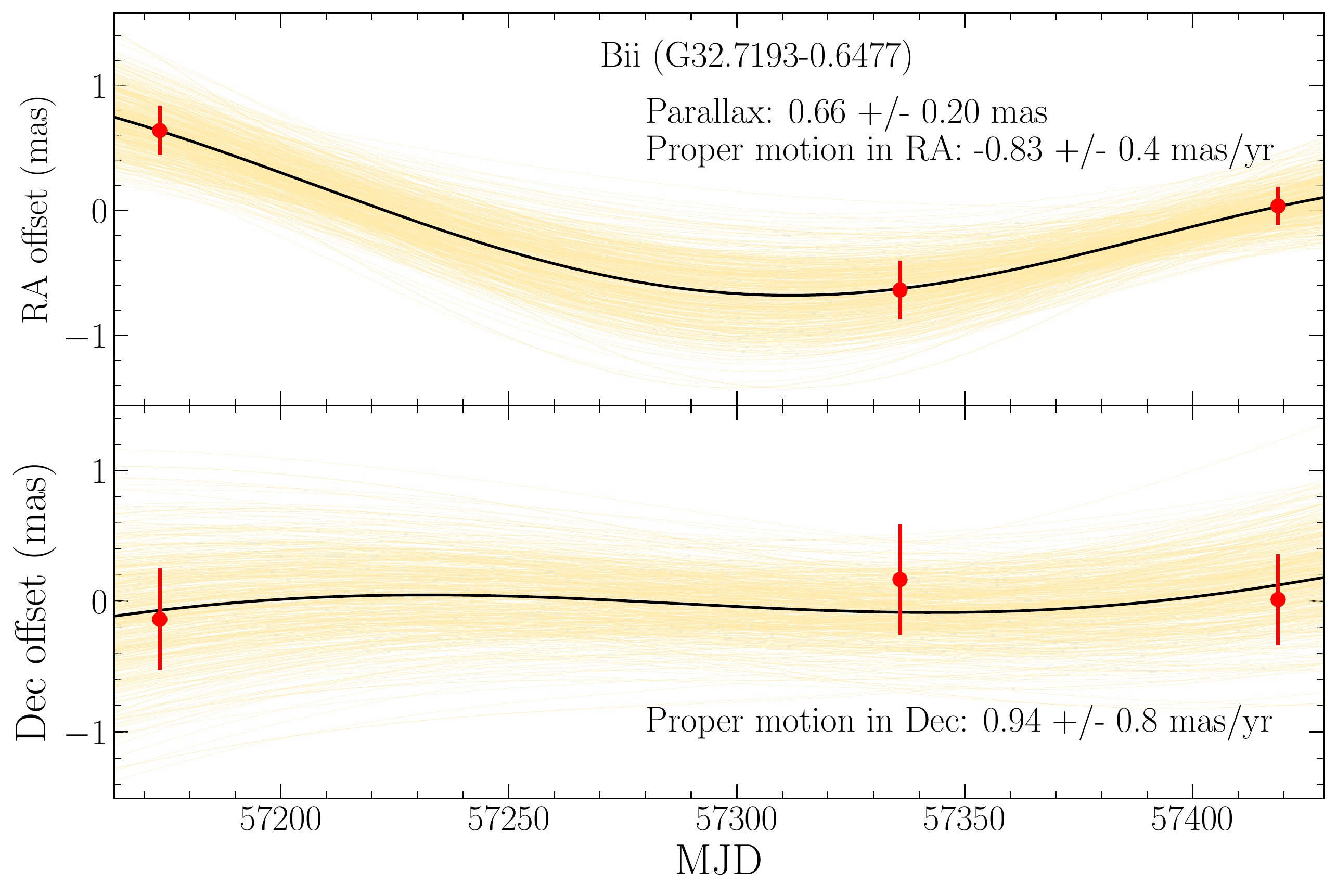}  \\
\includegraphics[width=9cm,height=7cm,angle=0]{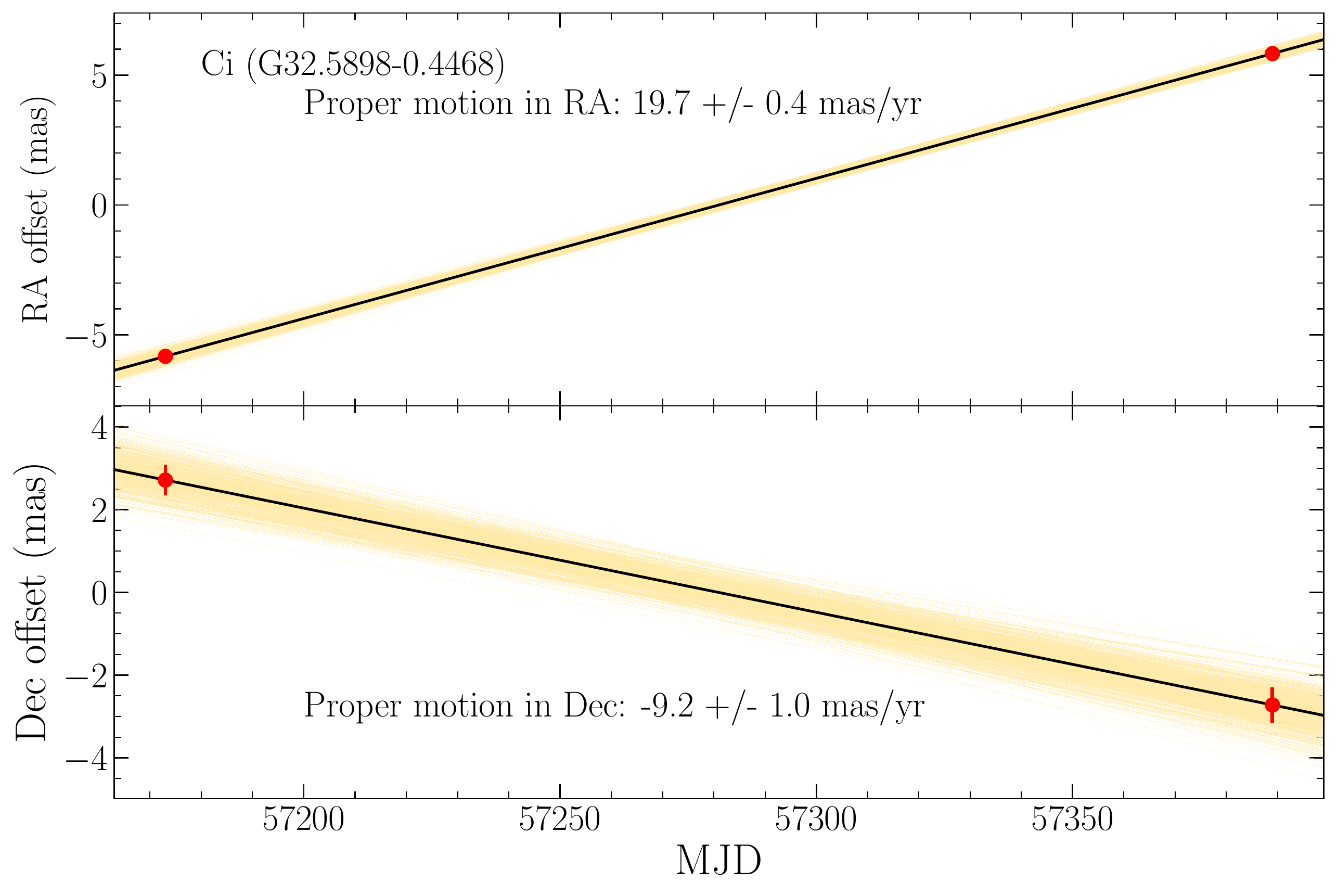} & 
\includegraphics[width=9cm,height=7cm,angle=0]{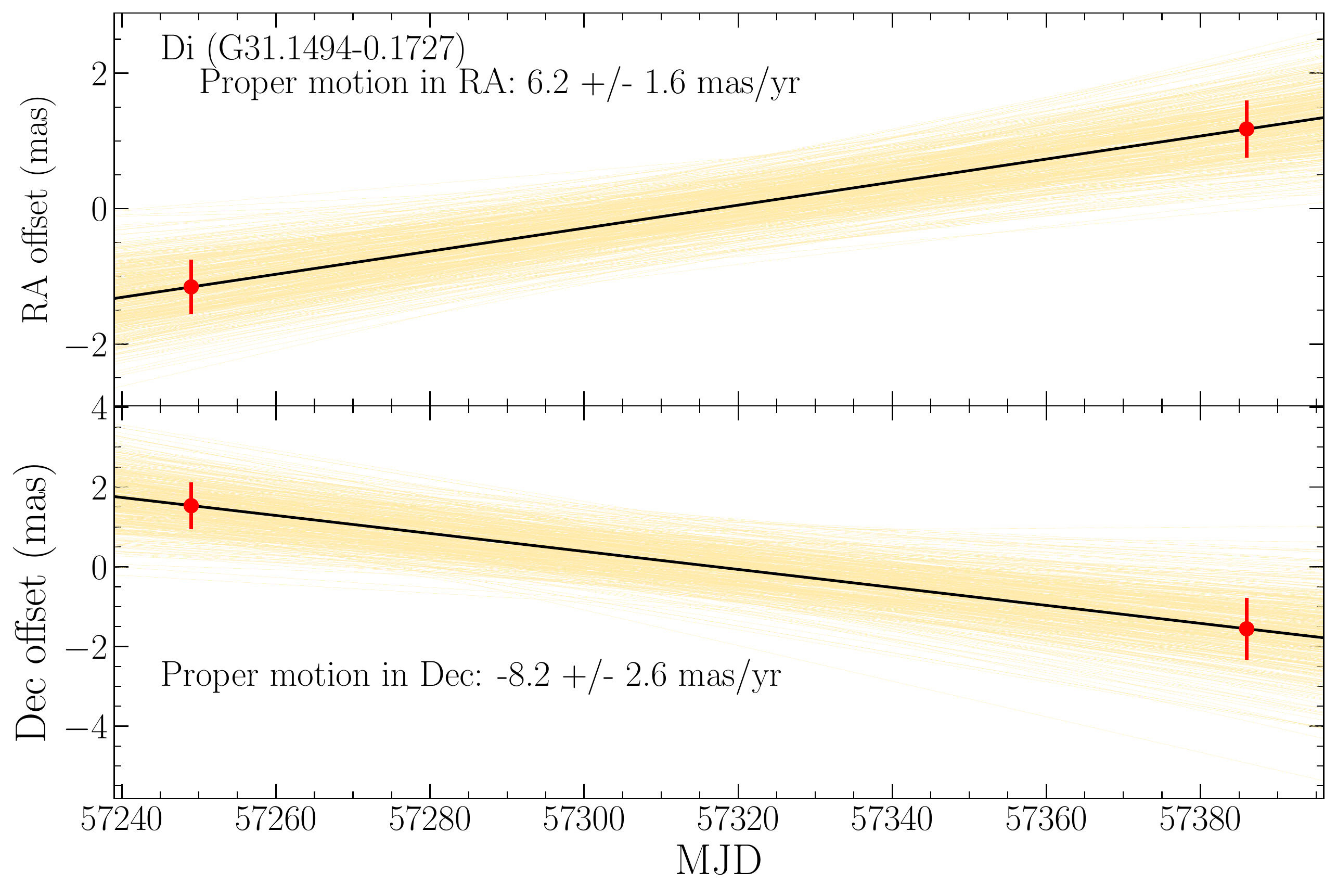} \\
\includegraphics[width=9cm,height=7cm,angle=0]{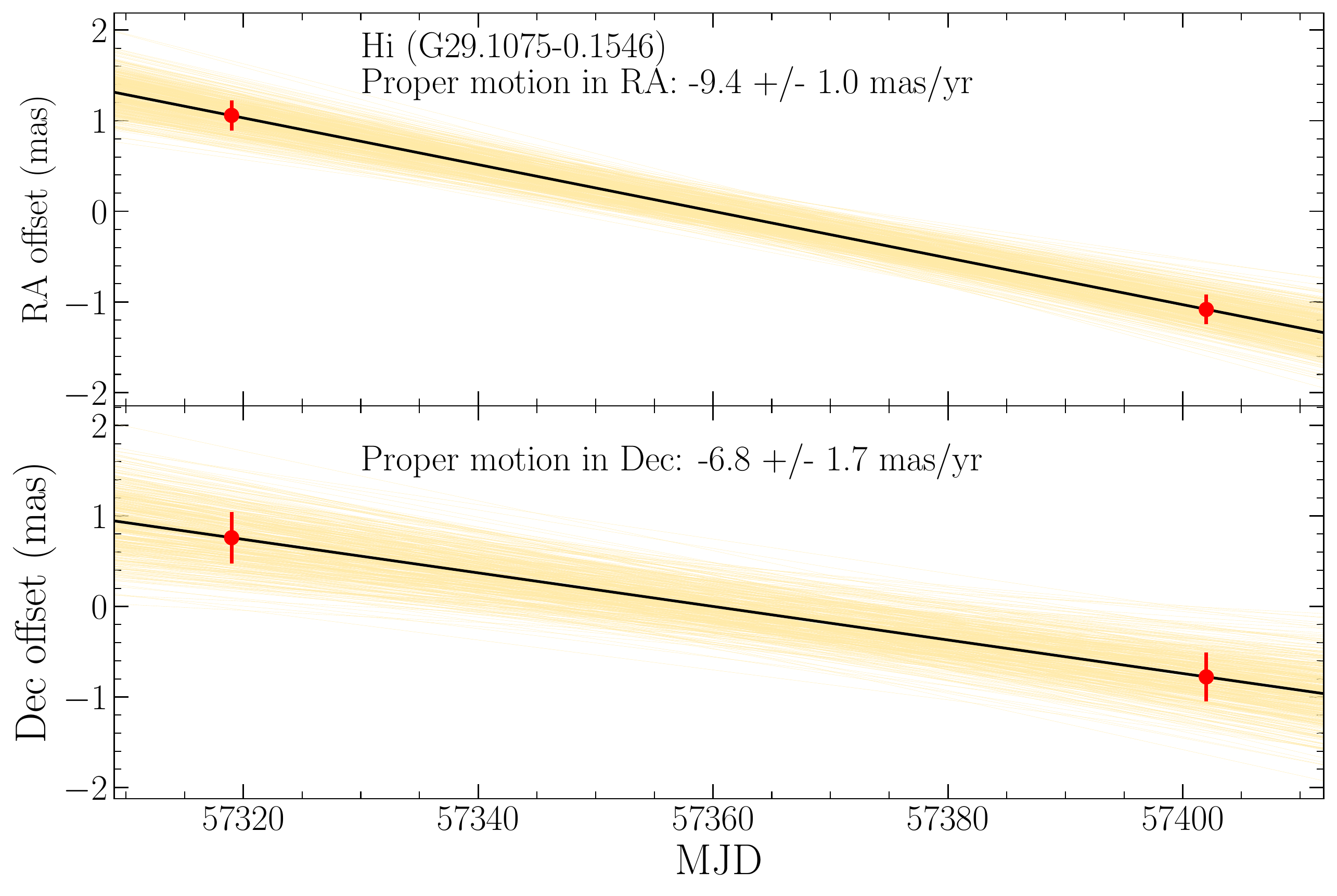} & 
\includegraphics[width=9cm,height=7cm,angle=0]{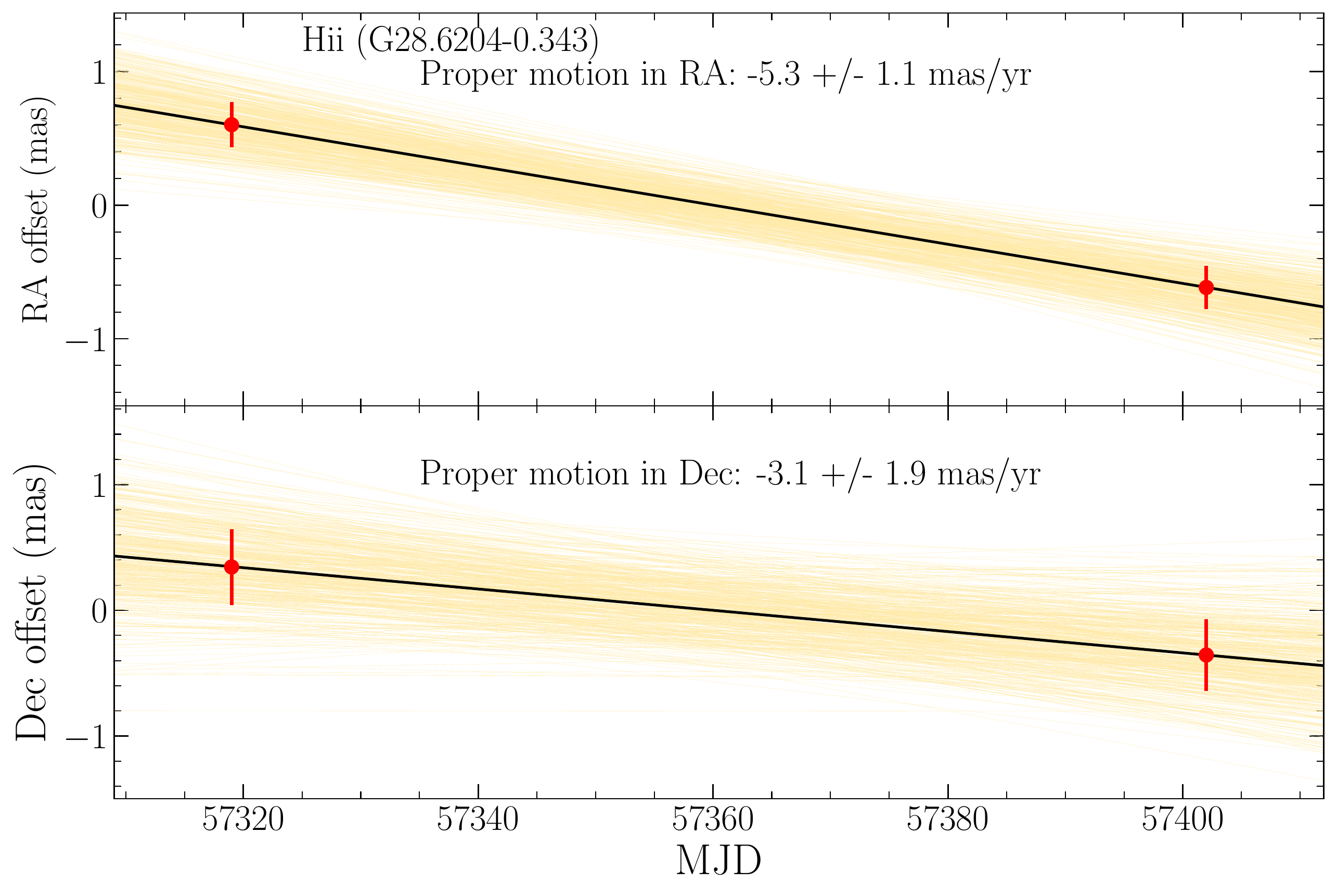} \\
\end{tabular}
\end{center}
\caption{ \label{timephase} Proper motion fits of A\romannumeral 1\,, C\romannumeral 1\,, D\romannumeral 1\,, H\romannumeral 1\, and H\romannumeral 2\,, and parallax and proper motion fits of B\romannumeral 2\,. The top and bottom panels in each figure represent the motion in the plane of the sky relative to the fitted reference position in RA and Dec, respectively. The offset of the target positions relative to the reference position is denoted by the red markers and is overlaid with the trajectory given by the best fitting astrometric trace (proper motion and parallax, where relevant) and is depicted by the black solid line. The yellow lines indicate the motions for 1000 randomly-selected draws from the posterior distribution.}
\label{motionfits1}
\end{figure*}

The PYMC3 python package \citep{Salvatier2016} was used to implement a Hamiltonian Markov Chain Monte Carlo \citep[MCMC,][]{Neal2012} technique with a No-U-Turn Sampler \citep[NUTS;][]{Hoffman2011}. We first performed inference for all targets using a flat parallax prior between 0.1--5\,mas, and a flat proper motion prior for RA and Dec between $-$100--100\,mas\,yr$^{-1}$. The choice of the parallax prior is to cover a wide range of distance from 200\,pc to 10\,kpc and the proper motion prior is chosen such that it significantly exceeds the range of known BHXB proper motions \citep{Atri2019,Brown2018}. The results of the fitting are presented in Table \ref{vlba_survey_pm} and Figure \ref{timephase}. Given the small sample size, we could not disentangle the contributions of parallax and proper motion for the sources with two epochs of observations and our inference indicated no significant difference between posterior and prior probability distributions for the parallax. This degeneracy also resulted in broadened posteriors for proper motion. Thus to verify whether these objects are moving or not, we performed inference using a simple model to fit a straight line to estimate the proper motion, assuming a zero parallax. In Table \ref{vlba_survey_pm} we report the posterior estimates for proper motion based on both models (with and without parallax), and use the results of a straight line proper motion fit in our further analysis (see Figure \ref{motionfits1}). There is a $>$1$\sigma$ discrepancy between the total proper motion of two targets, H\romannumeral 1\, and H\romannumeral 2, when fitting with and without a parallax prior. This is due to the short time baseline of only 80\,days between the two epochs of these two targets, as compared to the $>$130\,days for all other targets. \par
We could obtain a constraining parallax measurement with a $>$3$\sigma$ significance for only one target, B\romannumeral 2\phantom{1}(and with the caveat of the $\sim5.7\sigma$ detection in first epoch). We note that the errors on the proper motion fits of B\romannumeral 2\phantom{1}without the use of a parallax prior do not improve as compared to the fits with the parallax prior. There are two other targets that were detected on three epochs, E\romannumeral 1\, and  E\romannumeral 2\,, are detected as extended structures. Parallax and proper motion measurements of extended sources are not reliable as we cannot ascertain that the centroid detected in the images is of the same astrophysical region of the source in different epochs. Changing $uv$-coverage can also contribute to the shift in the peak brightness of resolved sources. Thus we do not fit for proper motion and parallax for these two sources. Amongst the other sources, we could measure the total proper motions of three targets, i.e., H\romannumeral 1, D\romannumeral 1\phantom{1}and C\romannumeral 1, with a $>$3$\sigma$ significance, of one target (H\romannumeral 2) with just below a 3$\sigma$ significance and of A\romannumeral 1\phantom{1}with a 2$\sigma$ significance. We could also use the sizes of the resolved targets and the size upper limits for the compact sources to estimate their observed brightness temperature. A brightness temperature of $>$10$^{6}$\,K estimated using VLBI is usually indicative of a non-thermal synchrotron emission mechanism for the radio emission. We use $T_{\mathrm{b}}=1.22 \times 10^{12} \frac{S_{\mathrm{tot}}}{ \theta_{\mathrm{maj}}\theta_{\mathrm{min}}\nu^{2}} \mathrm{K}$ to estimate the brightness temperature, where $S_{\mathrm{tot}}$ is the total fitted flux density of the source, $\theta_{\mathrm{maj}}$ and $\theta_{\mathrm{min}}$ are the Gaussian fits to the major and minor axes to the source size in milliarcseconds, and $\nu$ is the observing frequency in units of GHz. The results are summarised in Table \ref{tb_spec}.

\begin{table*} 
  \centering
  \caption{Results of the MCMC fitting code to determine proper motions and parallaxes of the compact targets detected at two or more epochs. The targets are arranged in decreasing order of the significance of proper motion measurement, where significance is defined as the ratio between the estimated value and its uncertainty. See Figure \ref{motionfits1} for plots.}
    \begin{tabular}{lccccc}
     Target & $\pi$ & $\upmu_{\alpha}\cos\delta$ & $\upmu_{\delta}$ & $\upmu_{\rm total}$ & No. of\\
      & (mas) & (mas\,yr$^{-1}$) & (mas\,yr$^{-1}$) & (mas\,yr$^{-1}$)  & epochs\\
     \hline
     \multicolumn{6}{c}{Proper motion fit only } \\ \hline
    G32.5898-0.4468 (C\romannumeral 1) & -- & 19.7 $\pm$ 0.4 & -9.2  $\pm$ 1.0 & 22.0 $\pm$ 0.7  & 2\\
    G29.1075-0.1546 (H\romannumeral 1) & -- & -9.4 $\pm$ 1.0 & -6.8  $\pm$ 1.7 & 12.0 $\pm$ 2.0  & 2\\
    G31.1494-0.1727 (D\romannumeral 1) & -- & \phantom{1}6.2 $\pm$ 1.6 & -8.2 $\pm$ 2.6 & 10.0 $\pm$ 3.0 & 2 \\
    G28.6204-0.3436 (H\romannumeral 2) & -- & -5.3 $\pm$ 1.1 & -3.1  $\pm$ 1.9 & \phantom{1}6.2 $\pm$ 2.2  & 2\\
    G39.1105-0.0160 (A\romannumeral 1)  & -- & \phantom{1}2.3 $\pm$ 0.7 & -0.4 $\pm$ 1.1 & \phantom{1}2.4 $\pm$ 1.2  & 2\\
    G32.7193-0.6477 (B\romannumeral 2) & -- & -1.1 $\pm$ 0.4 & \phantom{1}0.4 $\pm$ 0.8 & \phantom{1}1.2 $\pm$ 0.8  & 3\\
     
     \hline
     \multicolumn{6}{c}{Proper motion and parallax fit (parallax prior 0.1-5 mas)} \\
     \hline
    G32.5898-0.4468 (C\romannumeral 1) & 0.1 -- 5.0       & \phantom{1}22.2 $\pm$ 1.4 & -6.0 $\pm$ 2.0 & 23.0 $\pm$ 2.4  & 2\\
    G29.1075-0.1546 (H\romannumeral 1) & 0.1 -- 5.0       & -22.0 $\pm$ 8.0 & -5.4 $\pm$ 1.9  & 23.0 $\pm$ 8.0 & 2\\
    G31.1494-0.1727 (D\romannumeral 1) & 0.1 -- 5.0       & \phantom{11}2.2 $\pm$ 2.7 & -4.3 $\pm$ 3.4 & \phantom{1}4.8 $\pm$ 4.4  & 2\\
    G28.6204-0.3436 (H\romannumeral 2) & 0.1 -- 5.0       & -19.0 $\pm$ 7.4 & -1.7 $\pm$ 2.0 & 19.0 $\pm$ 7.7 & 2\\
    G39.1105-0.0160 (A\romannumeral 1) & 0.1 -- 5.0       & \phantom{1}-0.6 $\pm$ 1.8 & \phantom{1}5.3 $\pm$ 3.3 & \phantom{1}5.3 $\pm$ 3.8  & 2\\
    G32.7193-0.6477 (B\romannumeral 2) & 0.66 $\pm$ 0.20 & \phantom{1}-0.8 $\pm$ 0.4 & \phantom{1}0.9 $\pm$ 0.8 & \phantom{1}1.4 $\pm$ 1.0  & 3\\
     \hline

    \end{tabular}%
    \label{vlba_survey_pm}
\end{table*}%

\section{Discussion}\label{Section 6}

\subsection{Source classification on the basis of proper motion}
Below, we have summarised the classifications of the eight targets that we detected on multiple epochs based on the new VLBA information and any potential multiwavelength counterpart (see also Table \ref{optical_radio}). We report the classification of six targets amongst these eight based on their proper motion and parallax measurement. Five out of the six targets appear to be Galactic, with the total proper motion of three targets being inconsistent with zero at a $\geq$3$\sigma$ level, of one target being inconsistent with zero at just below the 3$\sigma$ level, and the $>$3$\sigma$ parallax measurement of one target placing it within the Galaxy (see discussion below for B\romannumeral 2). The one remaining target (A\romannumeral 1) could not be conclusively classified as either Galactic or extragalactic with just a 2$\sigma$ significance proper motion measurement. Higher accuracy on the total proper motion measurement of this target will help conclude if they are Galactic or extragalactic. For the two resolved sources (E\romannumeral 1\, and E\romannumeral 2) that were detected on multiple epochs, the limited $uv$-coverage hindered the measurement of any motion between the centroid of the target between epochs. \par
Multiwavelength studies of the objects in our sample can help identify the nature of these sources. We performed a basic query on Vizier\footnote{https://vizier.u-strasbg.fr/viz-bin/VizieR} to look for possible multiwavelength counterparts for each target. The VLBA position measurements of our targets provide mas level accuracy but the catalogs we searched have larger uncertainties, especially towards the Galactic bulge and Plane. Thus, we used a conservative search radius of 1\,arcsecond and assessed each possible counterpart in this search radius on a case-by-case basis. Below, we report any possible counterparts in optical wavelengths from the Panoramic Survey Telescope and Rapid Response System \citep[Pan-STARRS;][]{Chambers2016} catalogue that has a $g$-magnitude limit of 23.2 and a positional accuracy of 0.01$^{\prime\prime}$ \citep{Chambers2016}, and the \textit{Gaia}-DR2 survey \citep[][]{Brown2018} that has a $g$-magnitude limit ranging between 18.2--21.3 \citep{Boubert2020} and a positional accuracy of 0.002$^{\prime\prime}$. We search for infrared counterparts in the UKIRT Infrared Deep Sky Survey \citep[UKIDSS;][]{Lawrence2007,Lucas2008} that has a $K$-magnitude limit of 23 and a positional accuracy of 0.1$^{\prime\prime}$ \citep{Lucas2008}. \par
\begin{figure}
\centering
\includegraphics[width=0.5\textwidth]{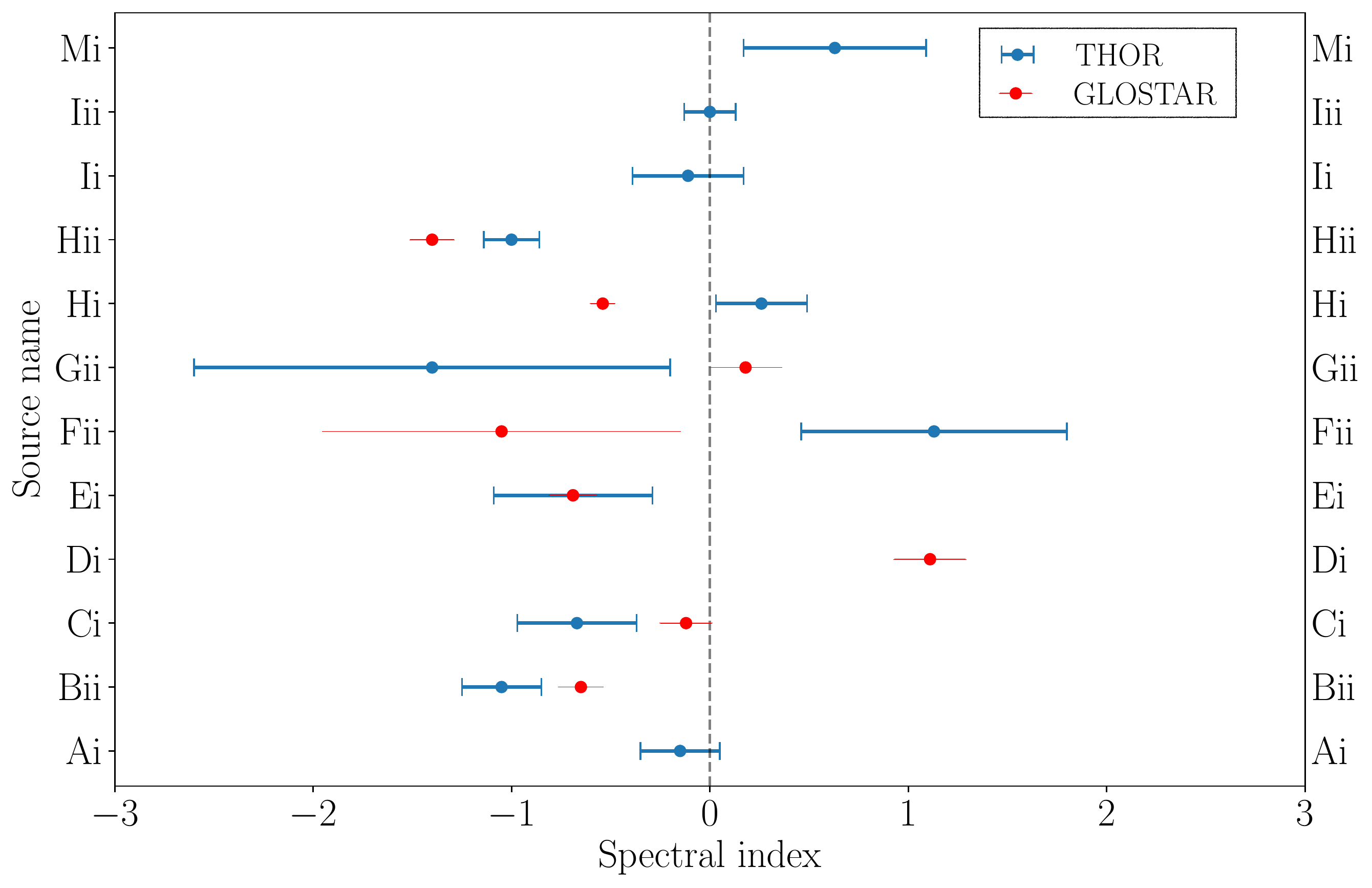}
\caption{Comparison between the spectral indices reported in the THOR survey (1.06--1.95\,GHz) that are depicted by blue markers, and in the GLOSTAR survey (4.7--6.9\,GHz.) that are depicted by red markers. The grey dashed line is marked at spectral index 0 to. 8 targets had both a THOR and a GLOSTAR spectral index measurement, 5 targets only had a THOR spectral index reported and 1 target had only the GLOSTAR spectral index. 8 out of 14 targets are consistent with a flat spectrum within 1$\sigma$, 3 targets are consistent with a flat spectrum within 2$\sigma$ and the remaining 3 targets have a spectral index inconsistent with zero at a $>$2$\sigma$. }
\label{spec_plot}
\end{figure}

The sources in our VLBA survey were selected as they were flat spectrum sources in the B10 study. We cross-match our sources with The HI/OH/Recombination line survey of the Milky Way \citep[THOR;][]{Beuther2016,Bihr2016,Wang2018} and the Global View of Star Formation in the Milky Way \citep[GLOSTAR;][]{Medina2019} to obtain additional information about the spectral indices of any matched sources. Both THOR and GLOSTAR surveys are being conducted by the VLA with a resolution of 10--25$^{\prime\prime}$ and 18$^{\prime\prime}$, respectively. The THOR spectral indices are based on flux measurements in the frequency band 1.06--1.95\,GHz, and the GLOSTAR spectral indices cover the frequency band 4.7--6.9\,GHz. Drastically different spectral indices in these two bands could indicate a spectral break. However, as the two surveys are not simultaneous in time, have different resolutions based on the configurations they were observed in, and the sources are highly variable, we are not confident about the spectral index comparisons between the two surveys. On the other hand, consistency between the spectral indices from the THOR and GLOSTAR survey suggests that the spectral index measurements are likely correct (see Figure \ref{spec_plot}). We found spectral index measurements for the 14 sources we detected in our VLBA data, in either one or both the THOR and the GLOSTAR survey. As can be seen in Table \ref{tb_spec} and Figure \ref{spec_plot}, 8 out of the 14 sources are consistent with being a flat spectrum source within 1$\sigma$, which also agrees with the evaluation presented in B10. Three targets are clearly inconsistent with being a flat spectrum source in both the THOR and the GLOSTAR bands, and are discussed on a case-by-case basis below. \par
We also compare the classification we make of the targets based on proper motions to the source type suggested in the the Co-ordinated Radio and Infrared Survey for High Mass Star Formation (CORNISH) survey \citep{Purcell2008,Purcell2013}, a radio survey at 5\,GHz using the VLA. The CORNISH survey's online catalogue\footnote{https://cornish.leeds.ac.uk/public/catalogue.php} included cross-matches with UKIDSS and the Galactic Legacy Infrared Mid-Plane Survey Extraordinaire \citep[GLIMPSE;][]{Benjamin2003}, which helped in studying the spectral energy distribution of the targets and suggesting possible source types for the targets. \par
\subsubsection{Galactic sources}
\textbf{G32.7193-0.6477 (B\romannumeral 2)} was detected in three epochs and we were able to obtain loose constraints on the parallax of the target (0.66$\pm$0.20\,mas), see Figure \ref{motionfits1} and Table \ref{vlba_survey_pm}. The proper motion of the source is consistent with zero motion within 2$\sigma$, but the parallax fit  suggests that the target is possibly a Galactic radio source with slow projected movement. The target was not detected in various optical, X-ray and infrared catalogues in 2010 as searched by B10. We also could not find any multiwavelength counterpart of B\romannumeral 2. We use the GLOSTAR spectral index of $-$0.65$\pm$0.11 and extrapolate the VLA flux density of this target from 5\,GHz to 8.3\,GHz. The expected flux density at 8.3\,GHz is 0.7\,mJy\,beam$^{-1}$, which is consistent with our detected flux density of B\romannumeral 2\, in the VLBA observations. Since these targets are also highly variable, it is also possible that the target varied to a sub-mJy level. The CORNISH catalogue suggests that this target is a radio galaxy, which contradicts our suggestion that the source is Galactic in nature (see Table \ref{optical_radio}). As this source was seen as a compact source in our VLBA images, we estimated the lower limit on the observed brightness temperature of the source in the three epochs as $>$(0.7--1.2)$\times$10$^{6}$\,K, which could suggest a non-thermal emission mechanism for the observed radio emission. We used the data and pipeline provided by the UK Swift Science Data Centre at the University of Leicester to find all X-ray observations covering the target \citep{Evans2007} and estimate the 3$\sigma$ upper-limit on count rate to be 1.88$\times$10$^{-2}$\,ct\,s$^{-1}$ in the 0.3-10\,keV band. This translates to an upper limit of 1.5$\times$10$^{-12}$\,erg\,s$^{-1}$\,cm$^{-2}$ of unabsorbed flux (0.3-10\,keV) assuming a distance of 1.5\,kpc, hydrogen column density of 6.1$\times$10$^{21}$\,cm$^{-2}$ in the direction of B\romannumeral 2\, \citep{Green2019,Bahramian2015} and a photon index of 2 \citep{Plotkin2017}. The flat to slightly-inverted spectra \citep{Fender2001}, compact and Galactic nature makes G32.7193-0.6477 a promising candidate for a quiescent BHXB. Deeper X-ray observations and stronger constraint on the distance will help in further investigation of this 
\par 

\textbf{G32.5898-0.4468 (C\romannumeral 1)} The proper motion of C\romannumeral 1\, is measured with a significance of 20$\sigma$, convincingly demonstrating that the source is a Galactic radio source. The source was not detected in any of the catalogues searched by B10. We found an optical and IR source in the Pan-STARRS catalogue and the UKIDSS catalogue, respectively, within 0.93$^{\prime\prime}$ from the VLBA position of C\romannumeral 1. Owing to its large proper motion, we estimated the position of the target at the mean epoch of the Pan-STARRS detection (MJD 56090) by extrapolating the measured proper motion of the target from the first epoch of the VLBA observation (MJD 57173). We find that this position is also 0.85$^{\prime\prime}$ offset from the Pan-STARRS position. The positional accuracy of 0.01$^{\prime\prime}$ of the Pan-STARRS source \citep{Chambers2016} and 0.1$^{\prime\prime}$ with UKIDSS \citep{Lucas2008} makes it an unlikely optical/IR counterpart of C\romannumeral 1. The GLOSTAR survey reports a spectral index of $-$0.12$\pm$0.13 for C\romannumeral 1, which is consistent with being a flat spectrum source. There is also a \textit{Gaia} source $0.88^{\prime\prime}$ away that has a proper motion of $-$4.4$\pm$3.1\,mas\,yr$^{-1}$ in the RA direction and $-$3.5$\pm$2.6\,mas\,yr$^{-1}$ in the Declination direction. The difference in the proper motion of the \textit{Gaia} source and the VLBA source, along with the high precision position measurement of \textit{Gaia} (1.5\,mas) makes the \textit{Gaia} source an unlikely optical counterpart of C\romannumeral 1. The lower limit on the observed brightness temperature of this source is (2.4--3.3)$\times$10$^7$K, which suggests a non-thermal emission origin for the radio emission. We used the data and pipeline provided by the UK Swift Science Data Centre at the University of Leicester to find all X-ray observations covering the target \citep{Evans2007} and estimate the 3$\sigma$ upper-limit on count rate to be 9.1$\times$10$^{-3}$\,ct\,s$^{-1}$ in the 0.3-10\,keV band. The distance to this source is unconstrained, so assuming a conservative distance of 1\,kpc, the count rate translates to an upper limit of 1.5$\times$10$^{-12}$\,erg\,s$^{-1}$\,cm$^{-2}$ of unabsorbed flux (0.3-10\,keV).

\textbf{G31.1494-0.1727 (D\romannumeral 1)}
This target was observed and successfully detected in two epochs, albeit with a significance of 5--6$\sigma$, enabling a proper motion measurement. The total proper motion of this source is significant at the 3$\sigma$ level (see Figure \ref{motionfits1} and Table \ref{vlba_survey_pm}), giving support to the Galactic nature of the source. We did not find any multiwavelength counterpart of D\romannumeral 1\, within 1$^{\prime\prime}$ and the B10 study also could not find emission in any other wavelength from this target. The CORNISH catalogue classified this source as a high-redshift radio galaxy that is undetected in infrared wavelengths \citep[IR Quiet;][]{Collier2014}, whereas our proper motion measurement suggests that this source is likely Galactic in nature.  \par

\textbf{G29.1075-0.1546 (H\romannumeral 1)} was observed and detected in two epochs. The high total proper motion of the target is significant at the 6$\sigma$ level (see Figure \ref{motionfits1} and Table \ref{vlba_survey_pm}), which is strong evidence that the target is a Galactic source. A UKIDSS target is found 0.75$^{\prime\prime}$ away from the radio position, but is unlikely an infrared counterpart to H\romannumeral 1\phantom{1}given that the nominal UKIDSS positional accuracy is 0.1$^{\prime\prime}$. THOR and GLOSTAR measured a spectral index of 0.3$\pm$0.2 and $-$0.5$\pm$0.1, respectively, which could indicate a spectral break between the two wavebands with the caveat that the spectral index measurements in both the surveys are not simultaneous. Although Vizier suggests that H\romannumeral 1\, is associated with an ultracompact HII region \citep{Molinari1998}, there is a 106$^{\prime\prime}$ separation between the VLA radio source and the infrared counterpart. According to \citet{Molinari1998} a stronger infrared association ($<$40$^{\prime\prime}$) is integral to the ultracompact HII classification of the source. We thus reject this classification of  H\romannumeral 1. Based on the upper limit on the compact size of the source, we estimate an observed brightness temperature of $>$(2.5--2.7)$\times$10$^7$K. The CORNISH catalogue suggested that this target is an IR-Quiet source (i.e a radio-loud but infrared-quiet galaxy), and an extragalactic identification is in conflict with our suggestion that the source is in the Galaxy.\par

\textbf{G28.6204-0.3436 (H\romannumeral 2)} was detected in two epochs and a proper motion with a 3$\sigma$ significance was measured (see Figure \ref{motionfits1} and Table \ref{vlba_survey_pm}). We suggest that the source is likely Galactic in nature. The nearest infrared UKIDSS target is 0.73$^{\prime\prime}$ away from the radio position, but the 0.1$^{\prime\prime}$ positional accuracy makes it an unlikely counterpart of H\romannumeral 2. The spectral index of this source is $-$1.0$\pm$0.1, as consistently reported in the THOR survey and the GLOSTAR survey, which is typical for a pulsar \citep{Jankowski2018}. As H\romannumeral 2\, is not a known pulsar, it is possible that it is a highly scintillating and a temporally scatter-broadened pulsar and is thus undetected in the low frequency pulsar surveys \citep{Maan2018}. If a pulsar has a wide pulse profile it can be missed by pulsar timing campaigns, and could be a reason why this target is not a known pulsar. Additionally, this target could be like MTP13, which is a radio loud but X-ray quiet magnetar candidate \citep{Caleb2022}. The CORNISH catalogue classifies this source as an IR-quiet radio galaxy due to the lack of any infrared counterpart of this source, which is inconsistent with our classification of the source as being Galactic in nature.\par

\subsubsection{Inconclusive}

\textbf{G39.1105-0.0160 (A\romannumeral 1)} was detected in two epochs in our survey and the total proper motion is only measured at a $<$2$\sigma$ significance level (See Figure \ref{motionfits1} and Table \ref{vlba_survey_pm}). B10 did not detect this target in the infrared, optical and X-ray catalogues available at the time, and a Vizier search in the current available catalogues also did not show any matches for the source at any other wavelength. The THOR survey estimates a preliminary spectral index of $-$0.15$\pm$0.1 for A\romannumeral 1, which is consistent with the source being a flat spectrum source. The target is detected as a compact source and the lower limit of the brightness temperature is $>$3$\times$10$^6$K. The CORNISH classification of this source is an IR-Quiet radio galaxy. Based on just the proper motion, A\romannumeral 1\,could be an extragalactic radio source, or a Galactic radio source with a slow projected speed.\par
\textbf{G30.1038+0.3984 (E\romannumeral 1)} This target was observed in 3 epochs spread over a year, and was detected in all three epochs. However, the target shows a resolved structure in all three epochs, and centroid measurements for resolved sources are not always consistent with the astrophysical core of the source. We do not have a good enough $uv$-coverage to sample the resolved structure of the target well enough to determine how the centroid of the target shifts. We will need longer duration observations to determine if the source is extragalactic or Galactic in nature. This target is classified as an IR-Quiet radio galaxy in the CORNISH catalogue. There is a PANSTARSS source 0.76$^{\prime\prime}$ away from the VLBA position of E\romannumeral 1, but since the positional accuracy of PANSTARSS is 0.01$^{\prime\prime}$ it is not a likely optical counterpart of E\romannumeral 1. The THOR and GLOSTAR spectral of the source are consistent. The brightness temperature of the source is in the range (1.1--1.6)$\times$10$^6$K. \par

\textbf{G29.7161-0.3178 (E\romannumeral 2)} This target was detected in 3 epochs, but we could not measure a proper motion of the source as it shows a resolved structure in the VLBA data. This suggests that the target is inconsistent with being a radio star, as was suggested in the CORNISH catalog. The brightness temperature of the source is in the range (0.9--1.1)$\times$10$^6$K. E\romannumeral 2\phantom{1}was not detected in any other wavelength in the original B10 study. In the current release of UKIDSS, we find an infrared source 0.75$^{\prime\prime}$ from the radio position of E\romannumeral 2, which is an unlikely counterpart of E\romannumeral 2.

\subsection{Non-detections and single-epoch detections}\label{non-detection-discussion}
We did not detect 19 targets in any epoch in which they were observed. The non-detection of these targets could be because the radio emission from the targets was resolved out on VLBA baselines, even on the shortest baselines. We report J\romannumeral 1\ as a non-detection, however, we could see a strong fringe pattern in the image of this target even when imaging with the shortest VLBA baselines. This indicates that the target is extended but the limited sensitivity and resolution of the shorter baselines of the VLBA resulted in a non-detection of the target. As is clear from the THOR and GLOSTAR spectral indices of some of the VLBA sources we have detected, it is possible that some of the sources have an inverted spectrum as opposed to the B10 estimation of these sources being flat-spectrum. This would decrease the expected flux density of the targets at 8.3\,GHz and be undetectable by VLBA filler-time observations. We have also found that some of the sources are getting resolved at VLBA baselines, even though they were seen to be compact sources at VLA baselines, making some faint and highly resolved sources undetectable in our VLBA filler-time observations. The limited $uv$-coverage of our VLBA observations could also lead to non-detection of faint, extended targets. The targets chosen in this study had a fractional variability of 150$\%$ and so it is also possible that the sources varied enough in flux density to push the targets to a level too faint to detect with our VLBA filler program. Thus, a combination of high radio variability, steep spectrum and extended structure could be reasons we did not detect 19 of the targets with the VLBA at 8.3\,GHz. \par
17 of the targets from our VLBA sample were also observed with the European VLBI Network (EVN) and the Westerbork Synthesis Radio Telescope (WSRT) in 2010 to study their spectra and structure by Y22. They detected all targets with WSRT but only detected M\romannumeral 1\, with the EVN, which we also detected with the VLBA. They compared the flux densities of the targets from the ASKAP survey RACS, the THOR survey and the VLASS \citep{Lacy2020}, and identified eight targets (A\romannumeral 2\,, J\romannumeral 1\,, K\romannumeral 1\,, L\romannumeral 3\,, N\romannumeral 1\,, N\romannumeral 2\,, O\romannumeral 1\, and O\romannumeral 2\,) that could have steep spectra. They do not detect any of these sources in their EVN observations, and it is thus unsurprising that we did not detect these sources in our VLBA observations as well. These targets are also suggested to be candidate radio AGN in the CORNISH catalogue \citep{Purcell2013}. O\romannumeral 1\, is also seen as an arcsecond-scale extended source in the VLASS and CORNISH survey. A few of the sources (K\romannumeral 2\,, M\romannumeral 2\, and L\romannumeral 3\,) that are undetected at mas level resolution in both the EVN and the VLBA data were seen to have a rising spectra by Y22 and are classified as radio AGN candidates by the CORNISH catalogue.  We did not detect J\romannumeral 2\,, which Y22 confirm is a young Galactic HII region and is an arcsecond-scale extended target, and thus was likely resolved out in our VLBA maps. \par
We obtained single epoch position measurements of six targets, one of which had problems in the correlation of the data in the second epoch. We find that the six targets (A\romannumeral 3\,, F\romannumeral 2\,, G\romannumeral 2\,, I\romannumeral 1\,, I\romannumeral 2\, and M\romannumeral 1) are detected as resolved structures. Four of these single epoch VLBA detections (A\romannumeral 3\,, I\romannumeral 1\,, I\romannumeral 2\, and M\romannumeral 1\,) were also observed by Y22, but Y22 only detected M\romannumeral 1\,. In the case of M\romannumeral 1, the extended structure of the source as seen in the VLBA agrees with the conclusion of Y22, wherein they suggest that this target is an intrinsically compact extragalactic source and appears resolved due to scatter broadening. The sources  A\romannumeral 3\,, I\romannumeral 1\,, I\romannumeral 2\, are seen in our VLBA data with a peak intensity of 4.7$\pm$0.1\,mJy\,beam$^{-1}$, 0.90$\pm$0.05\,mJy\,beam$^{-1}$ and 1.13$\pm$0.06\,mJy\,beam$^{-1}$, respectively. Comparing the peak intensity of the VLBA detections of these three targets with rms noise in the EVN image of M\romannumeral 1\,, it appears that amongst A\romannumeral 3\,, I\romannumeral 1\,, I\romannumeral 2\, only A\romannumeral 3\, was bright enough to have been detected by Y22, unless A\romannumeral 3\, varied below the detection threshold. Y22 also reports on the elongated structure of A\romannumeral 3\, as seen at 1.4\,GHz in the Multi-Array Galactic Plane Imaging Survey \citep[MAGPIS;][]{Helfand2006}, which is also in agreement with our resolved structure conclusion of this target. The sizes of the two images, however, differ. MAGPIS sees a 12.5\,arcsec-long extension towards South, whereas our VLBA data shows an extended structure towards the East of size 36$\pm$0.1\,mas and 8.3$\pm$0.5\,mas in the major and minor axis, respectively. It is possible that our high resolution VLBA map is resolving out the extended structure reported in MAGPIS. 

\subsection{Source classification on the basis of variability}
We detected eight targets with reliable position measurements over multiple epochs and these targets were detected in all epochs they were observed in. The measured flux densities of these targets at different epochs agreed with each other within 1$\sigma$, suggesting that they did not experience drastic variability between the epochs when they were observed. We did not detect 19 sources in any of the epochs they were observed in. B10 compared the fractional variability of their Galactic plane population with the extragalactic population of \citet{de2004} and arrived at the conclusion that six out of 39 targets with $\mathit{f}<$1.5 are background sources. They also suggested that within the remaining 33 targets that have high fractional variability ($\mathit{f}>$1.5), there are likely to be 5 more extragalactic sources. We find that at most 1 out of 6 targets whose proper motions we could measure could be extragalactic (if they are not slowly moving Galactic sources). Amongst the six sources that have only one effective position measurement, we are confident that M\romannumeral 1\, is an extragalactic source. Thus, we have identified two targets as extragalactic sources in the 14 VLBA detections ($\approx$14$\%$), which confirms the B10 suggestion of $\approx$15$\%$ extragalactic sources in the 33 highly variable sources. \par

\subsection{Optical extinction limitations}

\begin{table}
  \centering
  \caption{Distance limits, CORNISH catalogue classification and classification from this work. $d$ is the distance at which the extinction in the target's direction changes the apparent $V$ magnitude of the target by 3 magnitudes. CORNISH source is the source type suggested in the CORNISH catalogue by investigating the SEDs of the targets, wherein IR-Quiet is a infrared-quiet and radio-loud galaxy. The last column denotes the suggested nature of the source based on its proper motion measurement (in case of compact objects), i.e., Galactic or Inconclusive. We could not confidently say if the resolved targets were Galactic or extragalactic as astrometry of extended targets with limited $uv$-coverage is not reliable. }
    \begin{tabular}{lcll}
    \hline
     Target & d & CORNISH & This work  \\
            & (kpc)     & source  &            \\
     \hline 
    G39.1105-0.0160 (A\romannumeral 1)   & $>$0.97  & IR-Quiet      & Inconclusive\\
    G32.7193-0.6477 (B\romannumeral 2)   & $>$0.58  & Radio-galaxy  & Galactic\\
    G32.5898-0.4468 (C\romannumeral 1)   & $>$0.50  & --            & Galactic\\
    G31.1494-0.1727 (D\romannumeral 1)   & $>$0.76  & IR-Quiet      & Galactic\\
    G30.1038+0.3984 (E\romannumeral 1)   & $>$5.1   & IR-Quiet      & Inconclusive\\
    G29.7161-0.3178 (E\romannumeral 2)   & $>$0.96  & Radio star    & Inconclusive\\
    G29.1075-0.1546 (H\romannumeral 1)   & $>$0.99  & IR-Quiet      & Galactic\\
    G28.6204-0.3436 (H\romannumeral 2)   & $>$2.26  & IR-Quiet      & Galactic\\
     \hline
    
    \end{tabular}%
    \label{optical_radio}
\end{table}%
The search for optical and infrared counterparts in all sky surveys for the eight sources that we detected on multiple epochs, did not yield any good matches. The Galactic radio source discovered by \citet{Kirsten2014} could be classified as a possible BHXB candidate, VLA\,J2130+12, due to its flat to slightly inverted radio spectrum, an upper limit on the X-ray luminosity of the source, the detection of a possible low mass (0.1--0.2\,M$_{\odot}$) optical counterpart with an apparent magnitude of 24.87 $\pm$ 0.24 in the V band, using images from the \textit{Hubble Space Telescope} and the distance to the source. None of the objects identified in our survey as Galactic could be associated with an optical counterpart. The reason such an extensive study was possible for VLA\,J2130+12 was because the source was in the foreground of a very well studied globular cluster, M15. Also, VLA\,J2130+12 is at a height of $\sim$1\,kpc above the Galactic plane and thus it suffers from less extinction as compared to the sources in our study that are closer to the Galactic plane. Also, the apparent magnitude of the optical counterpart of VLA\,J2130+12 is below the sensitivity of all known optical surveys. The sample studied in this work was a selection from a Galactic plane survey and covered a region of 22$^{\circ}<l<$ 40$^{\circ}$ and $b\leq$0.7$^{\circ}$, where $l$ and $b$ are Galactic longitude and latitude, respectively. As a result, any possible optical and infrared emission from the targets is susceptible to high levels of extinction due to absorption and reddening from the interstellar dust. The non-detection of any optical counterparts in all of the current surveys suggest that either the target does not emit in optical/IR wavelengths, or that the optical/IR light from the target has suffered enough reddening to push the apparent magnitude of the target to beyond the detection capabilities of our optical/IR telescopes \citep[e.g.,][]{Schlafly2016}. The extent of reddening $E(g-r)$ is direction-dependent and is estimated as a function of $l$, $b$ and the distance to the source. The extinction effect on the V magnitude of the target is related to reddening by 3.2$E(g-r)$, i.e., the $V$ magnitude of a source will be dimmed by 3.2 magnitudes if the $(g-r)$ color is 1 magnitude redder than expected. \par
Since the distance to almost all of our targets is unknown, other than the loose constraints on the distance of B\romannumeral 2, it is not possible to derive the exact level of extinction that the $V$ band will suffer. A combination of a few all sky surveys like the Two Micron All Sky Survey \citep[2MASS;][]{Skrutskie2006}, Pan-STARRS and \textit{Gaia} has been used to make a 3D map\footnote{http://argonaut.skymaps.info/}\citep{Green2019} of the interstellar dust reddening  as a function of $l$, $b$ and the distance towards the target. Since we cannot estimate the exact amount of reddening, we decided to map how far the source can be for the extinction in the target's direction to change the apparent magnitude of the target by 3 magnitudes (see $d$ in Table \ref{optical_radio}). This gives an idea of the extreme adverse effect of extinction on the detection possibility of some of the targets (A\romannumeral 1, B\romannumeral 2, C\romannumeral 1, D\romannumeral 1, E\romannumeral 2\phantom{1}and H\romannumeral 1) even if they are very close, within 1\,kpc. The other targets (E\romannumeral 1\,and H\romannumeral 2) show the same amount of extinction although for a much larger distance, suggesting that there is a higher probability of detecting optical counterparts to these targets.  

\subsection{Population comparison}
\begin{figure}
\centering
\includegraphics[width=0.45\textwidth]{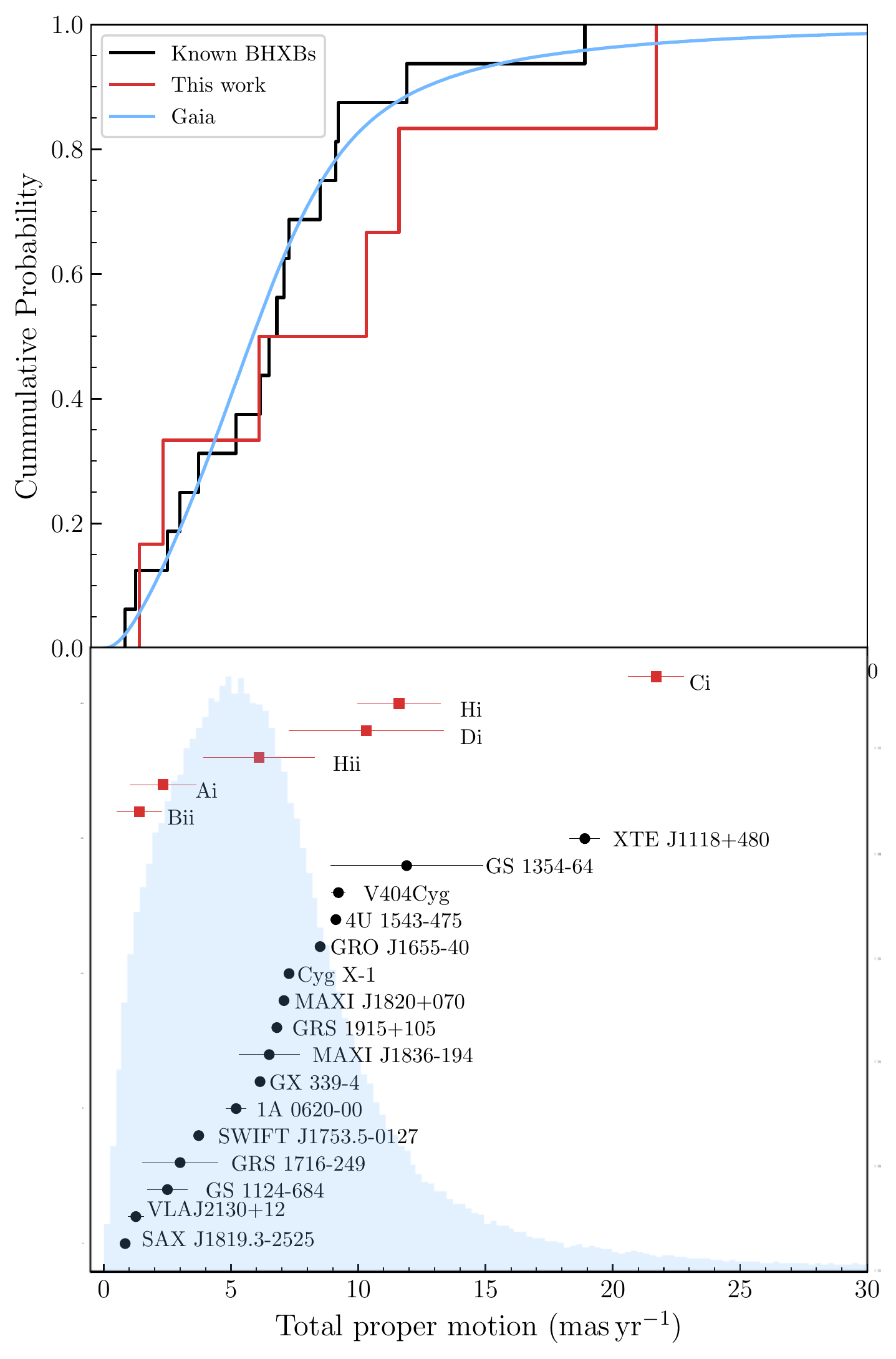}
\caption{Bottom panel - A forest plot of the total proper motion of the known BHXBs (black points), the sources for which proper motions were measured in this work (red points) and the \textit{Gaia} proper motions depicted with the blue histogram. Top panel - A cumulative distribution of the total proper motions of a sample of \textit{Gaia} sources (blue solid line), the radio sources in this work (red solid line) and the BHXBs known in our Galaxy (black solid line).}
\label{cdf_plot}
\end{figure}
The highly variable radio population in this work has at least five out of eight targets moving with respect to the ICRF, suggesting a Galactic origin. A search for radio stars by comparing the FIRST and the Sloan Digital Sky Survey (SDSS) revealed that only $\approx$1.2 out of a million optical stars in the the SDSS are seen as a bright ($>$1.25\,mJy) radio stars in the FIRST survey \citep{Kimball2009}. We tested the possibility that the radio sources in our sample with measured proper motions could be sampled from the same parent distribution as the sources for which the optical satellite \textit{Gaia} measured proper motions. We selected a random subset of 4,950,000 sources from the \textit{Gaia}-eDR3 \citep{Brown2021} dataset with magnitude $>$13 as the \textit{Gaia}-eDR3 sources $<$13 magnitude are suggested to have a different reference frame that might be offset from the ICRF \citep{Lindegren2020,Lindegren2020err}. We compared this \textit{Gaia}-eDR3 population with the six sources that we could measure proper motion for in our VLBA survey. We conducted the K-Sample Anderson-Darling test \citep{stephens1974,Scholz1987} with the null hypothesis that the two populations being considered belong to the same population, and performed the test using implementation in the R package \textsc{kSamples}\footnote{https://cran.r-project.org/web/packages/kSamples/index.html} of the test. We find that the null hypothesis probability is 22$\%$, which is inconclusive to determine whether the two samples come from a common population. We also tested the possibility that all targets with measured proper motions in our work and the BHXB population with known proper motions, as reported in \citet{Atri2019}, could be a part of the same parent population. The K-Sample Anderson-Darling test estimates the null hypothesis probability as 48$\%$, which is also inconclusive in regards to whether the two samples could belong to the same distribution. \par
Visual representation of the comparison of various populations tested above can be seen in Figure \ref{cdf_plot}. We plotted the total proper motions of the targets in this work and compared them to the total proper motions of the known BHXB population as reported in \citet{Atri2019} in Figure \ref{cdf_plot}, top panel. It is clear that the proper motions of both these samples lie in the same range, with one source (C\romannumeral 1) having a total proper motion greater than any BHXB known. This does not make the proper motion of C\romannumeral 1\, implausible as this could be a source that is closer than any known BHXB, and thus even a nominal space velocity projects to a high perceived proper motion. Due to lower number statistics the cumulative distribution of the total proper motions of the targets in this sample are indistinguishable from the sample selected from \textit{Gaia}-eDR3 and from the known BHXBs selected from \citet{Atri2019} (see Figure \ref{cdf_plot}, bottom panel).

\section{Conclusions}\label{Section 7}
With this project, we showed that measuring the proper motions of variable radio sources selected from wide field radio surveys is a viable technique to select a sample of compact Galactic targets and potentially search for quiescent BHXBs, or other classes of unknown radio transients. We observed 33 highly variable radio sources with the VLBA and detected 14 out of the 33 targets on one or more epochs. We could not find any optical or infrared counterparts for the sources we detected. \par
Eight targets were detected at multiple epochs and we could measure the proper motions for six of those targets. We can identify three targets (G32.5898-0.4468; C\romannumeral 1\,, G29.1075-0.1546; H\romannumeral 1\,, and G31.1494-0.1727; D\romannumeral 1\,) purely based on their high total proper motion. We find that G28.6204-0.3436; H\romannumeral 2\, is also likely Galactic as the proper motion is inconsistent with zero with a significance slightly lower than 3$\sigma$. We report that one target G32.7193-0.6477, B\romannumeral 2\, has a parallax measurement (0.66$\pm$0.20\,mas) that places it within the Galaxy. The proper motion of G39.1105-0.0160, A\romannumeral 1\, has large error bars and we could not conclusively identify if it is a Galactic or extragalactic source. We find that G30.1038+0.3984, E\romannumeral 1\, and G29.7161-0.3178, E\romannumeral 2\, have an extended structure. \par
We could identify six targets as having an extended structure and being resolved on VLBA baselines. Four of these targets G29.1978-0.1268, G\romannumeral 2; G27.8821+0.1834, I\romannumeral 1; G26.2818+0.2312, I\romannumeral 2\, and G23.6644-0.0372, M\romannumeral 1\, were imaged by concatenating data from different epochs, suggesting that the targets did not show significant enough motion between the epochs that would smear the image in the concatenated image. This could indicate that the targets are extragalactic, however, this is not enough information to classify the sources as extragalactic. We did not detect 19 targets in any epoch, which could be due to high radio variability of the sources, possibility of the sources having an extended structure and the poor $uv$-coverage of our VLBA observations or the sources having a steep spectral index. We will be following up the VLBA detections for more epochs to get stronger constraints on the proper motions and parallaxes of the sources, and will be conducting deep X-ray, optical and infrared observations to identify multiwavelength counterparts to the targets.
\section{Acknowledgements}\label{Section 8}
We would like to thank Jun Yang and Leonid Gurvits for sharing the results of their unpublished EVN observations of a few targets that overlapped with our dataset, which allowed us to improve the analysis of our VLBA data. We would also like to appreciate the efforts of the anonymous referees that helped improve this manuscript. This research was supported by Vici research program 'ARGO' with project number 639.043.815, financed by the Dutch Research Council (NWO). JCAM-J is the recipient of an Australian Research Council Future Fellowship (FT140101082), funded by the Australian government. BM acknowledges financial support from the State Agency for Research of the Spanish Ministry of Science and Innovation under grant PID2019-105510GB-C31 and through the Unit of Excellence Maria de Maeztu 2020-2023 award to the Institute of Cosmos Sciences (CEX2019-000918-M). COH is supported by NSERC Discovery Grant RGPIN-2016-04602. GRS is supported by NSERC Discovery Grants RGPIN-2016-06569 and RGPIN-2021-0400. The National Radio Astronomy Observatory is a facility of the National Science Foundation operated under cooperative agreement by Associated Universities, Inc. This work made use of the Swinburne University of Technology software correlator, developed as part of the Australian Major National Research Facilities Programme and operated under licence. This publication makes use of data products from the Two Micron All Sky Survey, which is a joint project of the University of Massachusetts and the Infrared Processing and Analysis Center/California Institute of Technology, funded by the National Aeronautics and Space Administration and the National Science Foundation.

\section{Data Availability}\label{Section 9}
All the VLBA data used in this paper is public and can be accessed from the NRAO data archive.




\bibliographystyle{mnras}
\bibliography{main} 





\bsp	
\label{lastpage}
\end{document}